\documentclass[12pt]{article}

\usepackage[pdftex]{graphicx}

\usepackage{amsmath} 
\usepackage{amssymb}
\usepackage{overpic}
\usepackage{fancybox}

\makeatletter
  
  \@addtoreset{equation}{section}
\makeatother

\begin{document}

%
\thispagestyle{empty}

\thisfancyput(123mm,-4mm){KOBE-COSMO-20-17}

\noindent
\textbf{\Large
Proof of new S-matrix formula from classical 
}

\vspace{4mm}

\noindent
\textbf{\Large 
solutions in open string field theory;}

\vspace{4mm}

\noindent
{\large 
or, Deriving on-shell open string field amplitudes without}

\vspace{2.4mm}

\noindent
{\large 
using Feynman rules, Part II }

\vspace{8mm}

\noindent
{
Masuda, Toru$^{a,\,}$\footnote[1]
{e-mail: masudatoru@gmail.com}
\quad 
Matsunaga, Hiroaki$^{a,\,b,\,}$\footnote[2]
{e-mail: matsunaga@math.cas.cz
}
\quad 
Noumi, Toshifumi$^{c,\,}$\footnote[3]
{e-mail: tnoumi@phys.sci.kobe-u.ac.jp
}

\vspace{8mm}

\noindent
$^a$CEICO, Institute of Physics, the Czech Academy of Sciences\\
\textit{Na Slovance 2, 18221 Prague 8, Czech Republic}\\

\noindent
$^b$Mathematical Institute, Faculty of Mathematics and Physics, \\
Charles University Prague\\
\textit{Sokolovska 83, Prague 3, Czech Republic}\\

\noindent
$^c$Department of Physics, Kobe University \\
\textit{1-1 Rokkodai-cho, Nada-ku, Kobe 657-8501, Japan}\\

\vspace{6mm}
\noindent
Abstract: 
We study relation between the gauge invariant quantity obtained in [arXiv:1908.09784] and the Feynman diagrams in the dressed $ \mathcal B_0 $ gauge in the open cubic string field theory.
We derive a set of recurrence relations that hold among the terms of this gauge invariant quantity.  
By using these relations, we prove that this gauge invariant quantity equals the $S$-matrix at the tree level.

We also present a proof that a set of new Feynman rules proposed in [arXiv:2003.05021 [hep-th]] reproduces the onshell disc amplitudes correctly by using the same combinatorial identities.  \\

\clearpage
\setcounter{page}{1}
\tableofcontents

\section{Introduction}

\noindent
(I) In a recent paper \cite{Masuda:2019rgv}, a new formula for the tree-level $S$-matrix in Witten's open string field theory  \cite{Witten} was proposed. 
In this formula, the tree-level $S$-matrix is expressed by using a classical solution $\Psi$ and a reference tachyon vacuum solution $\Psi_T$. 
Namely,
the formula reads
\begin{equation}
\label{eq:fml}
I_\Psi^{(N)}\equiv I_\Psi^{(N)}(\{\mathcal O_j\})
=C_N{\sum}' \int \prod_{j=1}^N (A+W_\Psi)\mathcal O_j,
\end{equation}
where 
\begin{itemize}
\item $\{\mathcal O_j\}$ $(1\le j\le N)$ is a set of states corresponding to 
incoming and outgoing open strings, which satisfies $Q_\Psi \mathcal O_j=0$. We call $\mathcal O_j$ an external state;
\item $N$ is the number of the external states; 
\item 
$C_N$ is a constant\footnote{$C_N$ will be determined in Section \ref{sec:section3.3}.} depending only on $N$; 
\item 
$A$ is given by $A=A_T-A_\Psi$, where $A_T$ is a so-called homotopy state for a reference tachyon vacuum solution $\Psi_T$, and 
$A_\Psi$ is a formal homotopy state for BRST operator around the classical solution  $Q_\Psi$; 
\item 
$W_\Psi$ is given by 
\begin{equation}
W_\Psi=A_T(\Psi-\Psi_T)+(\Psi-\Psi_T)A_T; 
\end{equation}
and
\item ${\sum}'$ represents the symmetrization over $\{\mathcal O_j\}$. We will give a precise characterization of ${\sum}'$ in Section 3. 
\end{itemize}

It is showed in \cite{Masuda:2019rgv} that $ I_\Psi^{(4)}$ matches 
the four-point $S$-matrix
with the boundary condition represented by $\Psi$,\footnote{We assumed that $\Psi$ is gauge equivalent to the Erler-Maccari solution \cite{Erler:2014eqa,Erler:2019fye}.}
 but the proof that $ I_\Psi^{(N)}$ matches the $S$-matrix for general $N$ has been missing. 
We will provide a proof of this statement in the present paper. 

Now, let us give an outline of our proof.
For convenience, let us call an expression of the following form an ``urchin" hereafter:
\begin{equation}
\int \prod_{j=1}^N X_j\mathcal O_j,
\end{equation}
where $X_i$ is either $A$ or $W_\Psi$. 
We find that, for a class of classical solutions (so-called the Okawa-type solutions), urchins are related to Feynman diagrams in the dressed $\mathcal B_0$ gauge. See formula \eqref{formula2}. 
We then derive a set of recurrence relations among the urchins \eqref{eq:formula3}. 
By using this relation, we characterize a vector space $V_N$ spanned by a set of linearly independent urchins. 
The proof that $I_\Psi^{(N)}$ is the $S$-matrix is obtained by converting the Feynman diagrams into a sum of urchins and expanding it in the basis of $V_N$. 
%
\\

\noindent
(II) Later, the authors of the paper \cite{Masuda:2019_2} proposed a new set of Feynman rules, which produces an expression for the $S$-matrix similar to $I^{(N)}_\Psi$. 
Let us denote this set of Feynman rules by $\mathcal R_\flat$. 
The propagator of $\mathcal R_\flat$ is given by
\begin{equation}
{\mathcal P}_\flat\phi=\frac{A}{2W_\Psi}\ast \phi+(-1)^{\text{gh}(\phi)} \phi\ast \frac{A}{2W_\Psi},
\end{equation}
and each Feynman diagram of $\mathcal R_\flat$ is an urchin. 
Indeed, for on-shell four-point amplitudes, 
$\mathcal R_\flat$ reproduces the very formula~\eqref{eq:fml};  
however,  for $N\ge 5$, the $S$-matrix calculated by using $\mathcal R_\flat$  
is  sum of urchins with non-trivial weights, and 
equivalence between this expression with $I_\Psi^{(N)}$ has not been shown so far. 
In this paper, we prove the equivalence 
 by using the above-mentioned recurrence relations among the urchins.   

The structure of this paper is as follows: 
 In Section 2, we present a formula for converting Feynman diagrams in the dressed $ \mathcal B_0 $ gauge into urchins. 
 In Section~\ref{sec:3.1}, using the results of Section~2, we derive a set of relations that hold among different urchins. 
Following some preliminary work in Section~\ref{sec:3.2}, we prove that the formula \eqref{eq:fml} represents the $S$-matrix in Section \ref{sec:section3.3}.  In Section~\ref{sec:4}, we prove that the expression obtained from the set of Feynman rules $\mathcal R_\flat$ matches the gauge invariant quantity $I_0^{(N)}$.
 In Section \ref{sec:conclusion}, we conclude the paper with some remarks. 

\section{Feynman rules in dressed $\mathcal B_0$ gauge}

We will give a brief review on the dressed $\mathcal{B}_0$ gauge in Section \ref{sec:2.1}, and then  we will present the formula linking  Feynman diagrams to urchins in Section~\ref{sec:2.2}.

\subsection{Summary of the dressed $\mathcal{B}_0$ gauge}
\label{sec:2.1}
The dressed $\mathcal{B}_0$ gauge was introduced by Erler and Schnabl~\cite{Erler:2009uj} as a generalization of the Schnabl gauge. The Schnabl-gauge condition is stated as
\begin{align}
\mathcal{B}_0\Psi=0\,,
\end{align}
where $\mathcal{B}_0$ is the zero mode of the $b$-ghost $b(z)$ in the sliver frame,
\begin{equation}
\mathcal B_0=\oint \frac{dz}{2\pi i}zb(z)\,.
\end{equation}
Its BRST transformation is the zero mode of the energy momentum tensor $T(z)$:
\begin{equation}
\{Q,\mathcal B_0\}=\mathcal{L}_0=\oint \frac{dz}{2\pi i}zT(z)\,.
\end{equation}
We also define
\begin{align}
\mathcal{B}_0^-=\mathcal{B}_0-\mathcal{B}_0^\star\,,
\quad
\mathcal{L}_0^-=\mathcal{L}_0-\mathcal{L}_0^\star\,,
\end{align}
which are derivatives with respect to star products. They commute with each other, $[\mathcal{B}_0^-,\mathcal{L}_0^-]=0$.
The operators, $\mathcal{B}_0^-$ and $\mathcal{L}_0^-$, act on $\{K, B, c\}$ as\footnote{Throughout the paper, we follow the convention of \cite{Masuda:2019rgv}. } 
\begin{align}
\nonumber
\frac{1}{2}\mathcal{B}_0^-K=B\,,\quad \frac{1}{2}\mathcal{L}_0^-K=K\,,
\\
\nonumber
\frac{1}{2}\mathcal{B}_0^-B=0\,,\quad \frac{1}{2}\mathcal{L}_0^-B=B\,,
\\
\nonumber
\frac{1}{2}\mathcal{B}_0^-c=0\,,\quad \frac{1}{2}\mathcal{L}_0^-c=-c\,.
\end{align}
Note that $\frac{1}{2}\mathcal{B}_0^-$ is trivial in the matter sector and $\frac{1}{2}\mathcal{L}_0^-$ simply counts the scaling dimension of the boundary operator inserted to the wedge state with zero width.

\subsubsection{Gauge fixing condition}

The dressed $\mathcal{B}_0$ gauge condition is stated as
\begin{align}
\label{dB_cond}
\mathcal{B}\Psi=0 
\end{align}
with $\mathcal{B}=\mathcal{B}_{F,G}$ defined by
\begin{align}
\mathcal{B}_{F,G}\,\bullet=\frac{1}{2}F(K)\mathcal{B}_0^-\left[F(K)^{-1}\bullet G(K)^{-1}\right]G(K)\,,
\end{align}
where $F(K)$ and $G(K)$ are  functions of $K$ satisfying certain regularity conditions. We also assume $F(0)G(0)=1$. 
If we take $F=G=e^{K/2}$, the gauge condition \eqref{dB_cond} reduces to the Schnabl gauge, whereas $F=G=1/\sqrt{1-K}$ corresponds to the gauge condition of the phantomless solution by Erler and Schnabl \cite{Erler:2009uj}}. 

The on-shell physical states in this gauge condition are of the form
\begin{equation}
\label{eq:phys1}
\varphi_i=F(K) O_i G(K)\,,
\end{equation}
where $O_i$ is an identity-based state of ghost number 1, satisfying $Q O_i=0$. More concretely, we may choose it as $ O_i=cV_i$, where $V_i$ is a dimension $1$ primary boundary operator in the matter sector inserted to the wedge state with zero width. Note that $O_i$ satisfies $[\mathcal{B}_0^-,O_i]=[\mathcal{L}_0^-,O_i]=0$.

\subsubsection{Propagator and simplified propagator}

The propagator in the dressed $\mathcal{B}_0$ gauge is given by
\begin{align}
\label{eq:propagator11}
{\mathcal P}_\text{d}={\mathcal P}_\text{s}Q {\mathcal P}_\text{s}^\star\,,
\end{align}
with\footnote{Here the subscript d stands for the dressed  $\mathcal{B}_0$ gauge, whereas the subscript s stands for the simplified propagator. }
\begin{align}
\label{eq:propagator2}
{\mathcal P}_\text{s}=
\lim_{\substack{\Lambda\to \infty\\\epsilon\to 0}}
\left(\int_0^\Lambda dse^{-s\mathcal{L}}+\int_\Lambda^{\Lambda+i\infty}ds e^{-s(\mathcal{L}-i\epsilon)}\right)\mathcal{B}\,.
\end{align}
Here 
$\mathcal L$ is the BRST transformation of $\mathcal{B}$:
\begin{align}
\mathcal{L}\,\bullet=\frac{1}{2}F(K)\mathcal{L}_0^-\left[F(K)^{-1}\bullet G(K)^{-1}\right]G(K)
\quad
{\rm with}
\quad
\mathcal{L}=\{Q,\mathcal{B}\}\,,
\end{align}
and $\Lambda$ and $\epsilon$ are regularization parameters to take care the on-shell pole, or in other words, the boundary term associated with the kernel of $\mathcal{L}$ (see, e.g.,~ \cite{Sen:2019jpm} for recent discussion). They are chosen such that $\Lambda\gg1$ and $\epsilon\Lambda\ll1$ . 

An important remark is that the dressed $\mathcal{B}_0$ gauge is a singular gauge just like the Schnabl gauge.  From the Feynman diagram point of view, it is essentially because the above formal propagator does not generate propagation of the open string midpoint. For a proper treatment of scattering amplitudes, we need to introduce a class of regular gauges as discussed in~\cite{Kiermaier:2007jg,Kiermaier:2008jy}. A conclusion there is that the naive propagator can be used for tree-level calculation, even though loop amplitudes have to be treated carefully with the regularized propagator in order to cover the closed string moduli appropriately.

In the following, we will assume $F(K)=G(K)$.

\subsubsection*{Conventions for Feynman diagrams}

We will use the following three maps to express the Feynman diagrams in formulas:
\begin{enumerate}
\item[i.] Star product $m: \mathcal H \otimes \mathcal H  \to \mathcal H$
\begin{equation}
m(\phi_i,\phi_j)=\phi_i\ast\phi_j,
\end{equation}
\item[ii.]   Propagator ${{\mathcal P}_\text{d}}$  in \eqref{eq:propagator11} (or the simplified propagator ${\mathcal P}_\text{s}$ in \eqref{eq:propagator2}),
\item[iii.] BPZ inner product $I: \mathcal H \otimes \mathcal H  \to \mathbb R$
\begin{equation}
I(\phi_i,\phi_j)=\int\phi_i\ast\phi_j.
\end{equation}
\end{enumerate}
We also define %
$
Y_x(\phi_i,\phi_j)={\mathcal P}_x \, m(\phi_i,\phi_j)
$ 
$(x=\text{d},\, \text{s})$
for notational simplicity.

\subsubsection*{$W$ and $A$ for the Okawa-type solutions}
\noindent
We will need $W(\equiv W_{\Psi=0})$ and $A_T$ to write down urchines.
When the tachyon vacuum solution  $ \Psi_T $ is the Okawa type 
\begin{equation}
\Psi_T=F(K)c\frac{KB}{1-F(K)^2}cF(K),
\end{equation}
$W$ and $A_T$ are given by
\begin{equation}
W=-F(K)^2,\quad A_T=\frac{1-F(K)^2}{K}B.
\end{equation}
Note that $A=A_T-A_{\Psi=0}$ is formally written as
\begin{equation}
A=-\frac{F(K)^2}{K}B=W\frac{B}{K}.
\end{equation}

%
%
%
%
%

\subsection{Converting Feynman diagrams into urchins}
\label{sec:2.2}
\subsubsection{Formula without the boundary term}
\noindent
Let us first present a formal argument, where the regularization of the propagator \eqref{eq:propagator2} is not took into account. 
We found the following simple relation:
\begin{equation}
\label{eq:noumieq}
{\mathcal P}_\text{d}
\left[ F(K)\mathcal O_i F(K)^2 \mathcal O_jF(K)\right]
=
F(K) \mathcal O_i\frac{1-F(K)^2}{K}B \mathcal O_jF(K),
\end{equation}
or in other words,
\begin{equation}
\label{eq:formula_firststep}
Y_\text{d}
(\varphi_i,\varphi_j)=\sqrt{-W} \mathcal O_iA_T \mathcal O_j\sqrt{-W}.
\end{equation}
Note that the left hand side of \eqref{eq:formula_firststep} is written in terms of the Feynman rules, and the right hand side 
is written with $W$, $A_T$, and $\mathcal O_j$, and 
can be regarded as a constituent part of the main part\footnote{
Here the main part of $I_\Psi^{(N)}$ refers to $I_\Psi^{(N)}$ excluding the contribution of $A_\Psi$.
See Section~4 of \cite{Masuda:2019rgv} for details. } of the formula \eqref{eq:fml}.\\

\noindent
\textbf{Proof of \eqref{eq:formula_firststep}:} 
Since $m(\varphi_i,\varphi_j)$ is BRST-exact and
\begin{equation}
{\mathcal P}_\text{d}={\mathcal P}_\text{s}-({\mathcal P}_\text{s}{\mathcal P}^\star_\text{s})Q,
\end{equation}
it holds that
\begin{equation}
Y_\text{d}(\varphi_k,\varphi_l)=Y_\text{s}(\varphi_k,\varphi_l). 
\end{equation}
We find that the right hand side $Y_\text{s}(\varphi_k,\varphi_l) $ is reduced to
\begin{align}
&
\frac{\mathcal{B}}{\mathcal{L}}\left[
F(K)
\mathcal O_kF(K)^2\mathcal O_lF(K)
\right]
\nonumber
\\
&=
\int_0^\infty ds 
F(K)
e^{-s\frac{1}{2}\mathcal{L}_0^-}
\frac{1}{2}\mathcal{B}_0^-
\left[\mathcal O_kF(K)^2\mathcal O_l\right]F(K)
\nonumber
\\
&=
-
\int_0^\infty dsF(K) e^{-s\frac{1}{2}\mathcal{L}_0^-}
\left[
\mathcal O_kH'(K)B\mathcal O_l\right]F(K)
\nonumber
\\
&=
-
\int_0^\infty ds F(K)\left[
\mathcal O_kH'(Ke^{-s})Be^{-s}\mathcal O_l\right]F(K)
\nonumber
\\
&=
F(K)
\left[
\mathcal O_kH(e^{-s}K)\frac{B}{K}\mathcal O_l\right]_{s=0}^{s=\infty}F(K)
\nonumber
\\
&=
F(K)\mathcal O_k(1-F(K)^2)\frac{B}{K}\mathcal O_lF(K)\,,
\end{align}
where $H(K)=F(K)^2$. $\square$\\

\noindent
By using the formula \eqref{eq:formula_firststep}, we can convert four-point Feynman diagrams into urchins or vice versa,
\begin{equation}
\label{eq:render1}
I(m( \varphi_1, \varphi_2),  Y_\text{d}(\varphi_3, \varphi_4))
= \int \mathcal O_1W\mathcal O_2W\mathcal O_3A_T\mathcal O_4W.
\end{equation}
By summing over $s$- and $t$-channels, the right hand side of \eqref{eq:render1} produces the main part of the gauge invariant quantity $I_0^{(N)}$.

\subsubsection{Formula with the boundary term}
We can reproduce the contribution of the boundary terms in $I_\Psi^{(N)}$ by more careful handling of the propagators. 
Let us consider the simplified propagator with regularization \eqref{eq:propagator2}. 
Acting this regularized propagator on the star product of two on-shell states in the dressed gauge gives
\begin{align}
\label{eq:formula1}
&{\mathcal P}_\text{s} \left[F\mathcal O_1H\mathcal O_2F\right]
\nonumber
\\
&=F\left[\int_0^\Lambda dse^{-\frac{s}{2}\mathcal{L}_0^-}\frac{1}{2}\mathcal{B}_0^-(\mathcal O_1H\mathcal O_2)\right]F
+F\left[\int_\Lambda^{\Lambda+i\infty} dse^{-s(\frac{1}{2}\mathcal{L}_0^--i\epsilon)}\frac{1}{2}\mathcal{B}_0^-(\mathcal O_1H\mathcal O_2)\right]F
\nonumber
\\
&=-F\left[\int_0^\Lambda dse^{-s}\mathcal O_1BH'(e^{-s}K)\mathcal O_2\right]F
-F\left[\int_\Lambda^{\Lambda+i\infty} ds
e^{i\epsilon s}
e^{-s}\mathcal O_1BH'(e^{-s}K)\mathcal O_2\right]F
\nonumber
\\
&=-F\mathcal O_1\left[\int_\lambda^1 dxH'(xK)
+
\frac{1}{K}H(\lambda K)\right]B\mathcal O_2F
+\mathcal{O}(\epsilon)
\,,
\end{align}
where we introduced $\lambda=e^{-\Lambda} \ll1$. Also in the last equality we used
\begin{align}
&\int_\Lambda^{\Lambda+i\infty} ds
e^{i\epsilon s}
e^{-s}H'(e^{-s}K)
\nonumber
\\
&=
\int_{e^{-(\Lambda+i\infty)}}^{\lambda} dx
x^{-i\epsilon}
H'(xK)
\nonumber
\\
&=
\left[
x^{-i\epsilon}
\frac{1}{K}H(xK)
\right]_{e^{-(\Lambda+i\infty)}}^\lambda
+i\epsilon\int_{e^{-(\Lambda+i\infty)}}^{\lambda} dx
x^{-1-i\epsilon}
\frac{1}{K}H(xK)
\nonumber
\\
&=
\frac{1}{K}H(\lambda K)
+\mathcal{O}(\epsilon)\,,
\end{align}
where we used $\lambda^\epsilon=e^{-\epsilon\Lambda}\simeq1$ and assumed that the integral in the third line does not give any $1/\epsilon$ singularity. 

The first term in the square bracket of the rightmost hand side of \eqref{eq:formula1} corresponds to $ A_T $, and the second term corresponds to $A_0= A_{\Psi=0} $,
\begin{equation}
-\lim_{\lambda\to 0}\int_\lambda^1 dxH'(xK)B
=
-\frac{1-F(K)^2}{K}B
=
A_T
\end{equation}
\begin{equation}
\lim_{\lambda\to 0}\frac{1}{K}H(\lambda K)B=\frac{B}{K}=A_{0}.
\end{equation}
Therefore, \eqref{eq:formula1} can be expressed as
\begin{equation}
\label{eq:formula}
Y_\text{d}(\varphi_i,\varphi_j)=
\sqrt{-W}
\mathcal O_iA\mathcal O_j\sqrt{-W}.
\end{equation}

\subsubsection{Generalization}
%
%
Now we wish to generalize \eqref{eq:formula} and present a formula for $n$ external states. 
Let us define $T_n(\varphi_1,...,\varphi_n)$ recursively by
\begin{equation}
\label{eq:defT00}
T_1(\varphi_1)=\varphi_1
\end{equation}
and
\begin{equation}
\label{eq:defT}
T_n(\varphi_1,...,\varphi_n)=\sum_{i=1}^{n-1} Y_\text{d}(T_i(\varphi_1,...,\varphi_i), T_{n-i}(\varphi_{i+1},...,\varphi_{n})). 
\end{equation}
The generalized formula is then given by
\begin{equation}
\label{formula2}
T_n(\varphi_1,...,\varphi_n)
=
\sqrt{-W}
\mathcal O_1A...\mathcal O_{n-1}A\mathcal O_n \sqrt{-W}.
\end{equation}
We can use this formula to convert a partial sum of Feynman diagrams into urchins. In the following, the arguments of $T_n$ may be omitted if there is no risk of confusion:
$T_n=T_n(\varphi_1,...,\varphi_n)$. 

\subsubsection{Genuine propagator and the simplified propagator}
\noindent
Note that we can use  the simplified propagator $\mathcal P_\text{s}$ instead of the genuine one 
 to calculate $T_n$
\begin{equation}
\label{eq:yeyyyey7yd4}
T_n=T_n^{(\text{s})},
\end{equation}
where $ T_n^{(\text{s})}$ is defined by
\begin{equation}
\label{eq:defT2}
 T_n^{(\text{s})}(\varphi_1,...,\varphi_n)=\sum_{i=1}^{n-1} Y_\text{s}(T_i^{(\text{s})}(\varphi_1,...,\varphi_i), T_{n-i}^{(\text{s})}(\varphi_{i+1},...,\varphi_n)),
\end{equation}
with the initial condition $T_1^{(\text{s})}=\varphi_1$. 
The formula \eqref{eq:yeyyyey7yd4} follows from the fact that 
${\mathcal P}^{-1}T_n$ 
is BRST closed %
\begin{equation}
\label{eq:cdyei3n}
Q{\mathcal P}^{-1}T_n(\varphi_1,...\,,\varphi_N)=0,
\end{equation}
where $\mathcal{P}^{-1}$ means amputation of the propagator attached to $T_n$, 
\begin{equation}
\mathcal{P}^{-1}T_n(\varphi_1,...\,,\varphi_n)=\sum_{i=1}^{n-1}m\left(T_i(\varphi_1,...\,,\varphi_i),T_{n-i}(\varphi_{i+1},...\,,\varphi_n)\right)\,.
\end{equation}
The statement \eqref{eq:cdyei3n} 
follows 
by induction on $n$.

%
\subsection{$T_n$ and scattering amplitudes}
\label{sec:ugr57u}
We can decompose $N$-point disc amplitudes into $(N-1)!$ pieces by the cyclic ordering of external states. For example, a four-point amplitude $\mathcal{A}^{(4)}$ can be decomposed as 
\begin{align}
\mathcal{A}^{(4)}=(\mathcal{A}_{1234}+\mathcal{A}_{1243}+\mathcal{A}_{1324}+\mathcal{A}_{1342}+\mathcal{A}_{1423}+\mathcal{A}_{1432})\,,
\end{align}
where $\mathcal{A}_{ijkl}$ has a definite cyclic ordering $[i,j,k,l]$. More generally, $N$-point amplitudes are decomposed as
\begin{align}
\mathcal{A}^{(N)}=\mathcal{A}_{12...N}+\text{ $\big((N-1)!-1\big)$ terms}.
\end{align}
We call amplitudes with a definite cyclic ordering ``color-ordered amplitudes" by analogy with non-Abelian gauge theories, and the cyclic ordering of the color-ordered amplitude is labeled by its subscripts as we have displayed.

By using $T_n$, the color-ordered amplitude  $\mathcal{A}_{12... N}$ can be written as
\begin{equation}
\label{eq:Npt_v1}
\mathcal{A}_{12... N}=I\big(T_1(\varphi_1),\mathcal{P}^{-1}T_{N-1}(\varphi_2,...\,,\varphi_N)\big)\,.
\end{equation}
We present a proof of \eqref{eq:Npt_v1} in Appendix \ref{sec:proofTnA} for completeness. 
Since we have the relation \eqref{eq:yeyyyey7yd4}, we can also use $T_{N-1}^{(\text{s})}$ in place of $T_{N-1}$. 
In this sense, we can calculate the on-shell amplitude using the simplified propagator, although it does not satisfy the BPZ property.

In fact, in addition to \eqref{eq:Npt_v1}, there are other formulas that represents $S$-matrices by using $T_n$. We will present them in Appendix \ref{sec:dbfd}. It gives a simple alternative proof that the formula \eqref{eq:fml} represents the $S$-matrix.
%

Let us conclude this section by presenting an example of 
turning the sum of Feynman diagrams into urchins.
\\

\noindent
\textbf{Example:} Consider the partial sum of the Feynman diagrams shown in Figure \ref{fig00}. This can be expressed as
\begin{equation}
\begin{split}
&I(T_3(\varphi_1,\varphi_2,\varphi_3), m(\varphi_4,\varphi_5))
=-\int \mathcal O_1 A \mathcal O_2 A \mathcal O_3 W \mathcal O_4 W \mathcal O_5 W. 
\end{split}
\end{equation}
The sum of these Feynman diagrams can also be expressed as 
\begin{equation}
\begin{split}
&I(\varphi_3, m(T_2(\varphi_4,\varphi_5), T_2(\varphi_1,\varphi_2)))+I(\varphi_1, m(T_2(\varphi_2,\varphi_3), T_2(\varphi_4,\varphi_5)))\\
=&-
\int \mathcal O_1 A \mathcal O_2 W \mathcal O_3 W \mathcal O_4 A \mathcal O_5 W
-\int \mathcal O_1 W \mathcal O_2 A \mathcal O_3 W \mathcal O_4 A \mathcal O_5 W.
\end{split}
\end{equation}
From these calculations, we find a relation among urchins
\begin{equation}
\begin{split}
&\int \mathcal O_1 A \mathcal O_2 A \mathcal O_3 W \mathcal O_4 W \mathcal O_5 W\\
=&\int \mathcal O_1 A \mathcal O_2 W \mathcal O_3 W \mathcal O_4 A \mathcal O_5 W
+\int \mathcal O_1 W \mathcal O_2 A \mathcal O_3 W \mathcal O_4 A \mathcal O_5 W.
\end{split}
\end{equation}
\begin{figure}[tbp]
\begin{center}
\begin{overpic}[width=11cm, bb=0 180 1024 608, 
]
{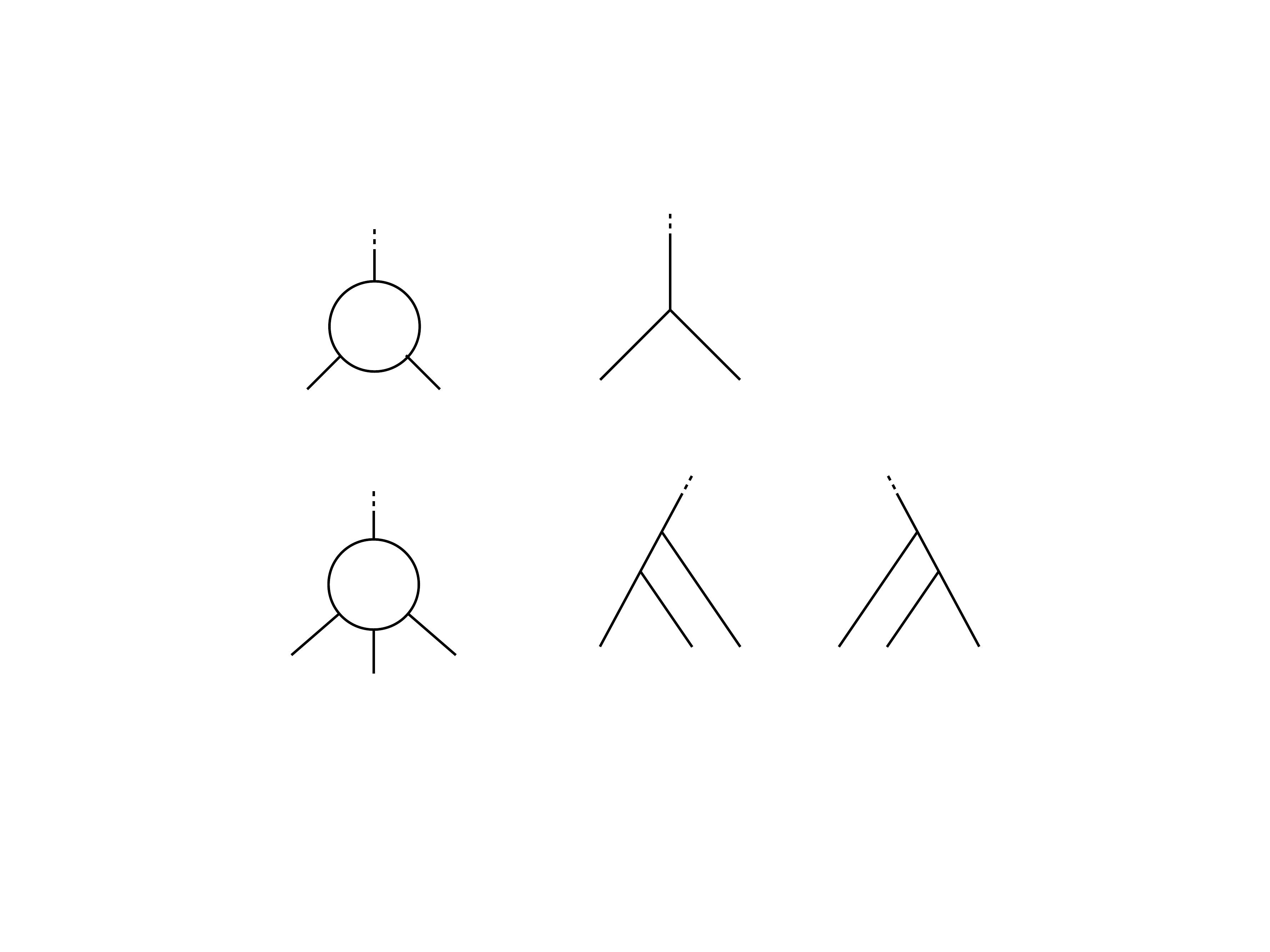}
\put(28.5,30.5){$2$}
\put(60,31){\small$\displaystyle = T_2(\varphi_i,\varphi_j)$}
\put(40,31){\small$\displaystyle = $}
\put(28.5,10){$3$}
\put(40,10){\small$\displaystyle = $}
\put(63,10){\small$\displaystyle +$}
\put(80,10){\small$\displaystyle = T_3(\varphi_i,\varphi_j,\varphi_k)$}
\put(22,24){\small$\displaystyle \varphi_i $}
\put(33,24){\small$\displaystyle \varphi_j $}
\put(46,24){\small$\displaystyle \varphi_i $}
\put(57,24){\small$\displaystyle \varphi_j $}
\put(21.5,3.2){\small$\displaystyle \varphi_i $}
\put(28.5,1.5){\small$\displaystyle \varphi_j $}
\put(35,3.2){\small$\displaystyle \varphi_k $}
\put(46,3.2){\small$\displaystyle \varphi_i $}
\put(51.5,3.2){\small$\displaystyle \varphi_j $}
\put(57,3.2){\small$\displaystyle \varphi_k $}
\put(65,3.2){\small$\displaystyle \varphi_i $}
\put(70.5,3.2){\small$\displaystyle \varphi_j $}
\put(76,3.2){\small$\displaystyle \varphi_k $}
\end{overpic}
\caption{A schematic picture for $T_2(\varphi_i,\varphi_j)$ and $T_3(\varphi_i,\varphi_j,\varphi_k)$.}
\end{center}
\end{figure}
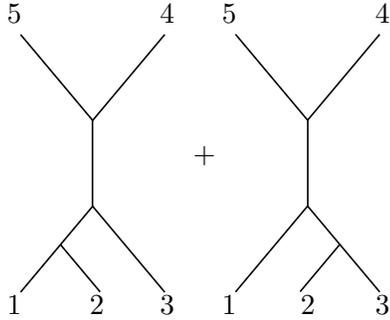
\begin{figure}[tbp]
\begin{center}
\begin{overpic}[width=11cm, 
bb=0 -30 1024 460 
]{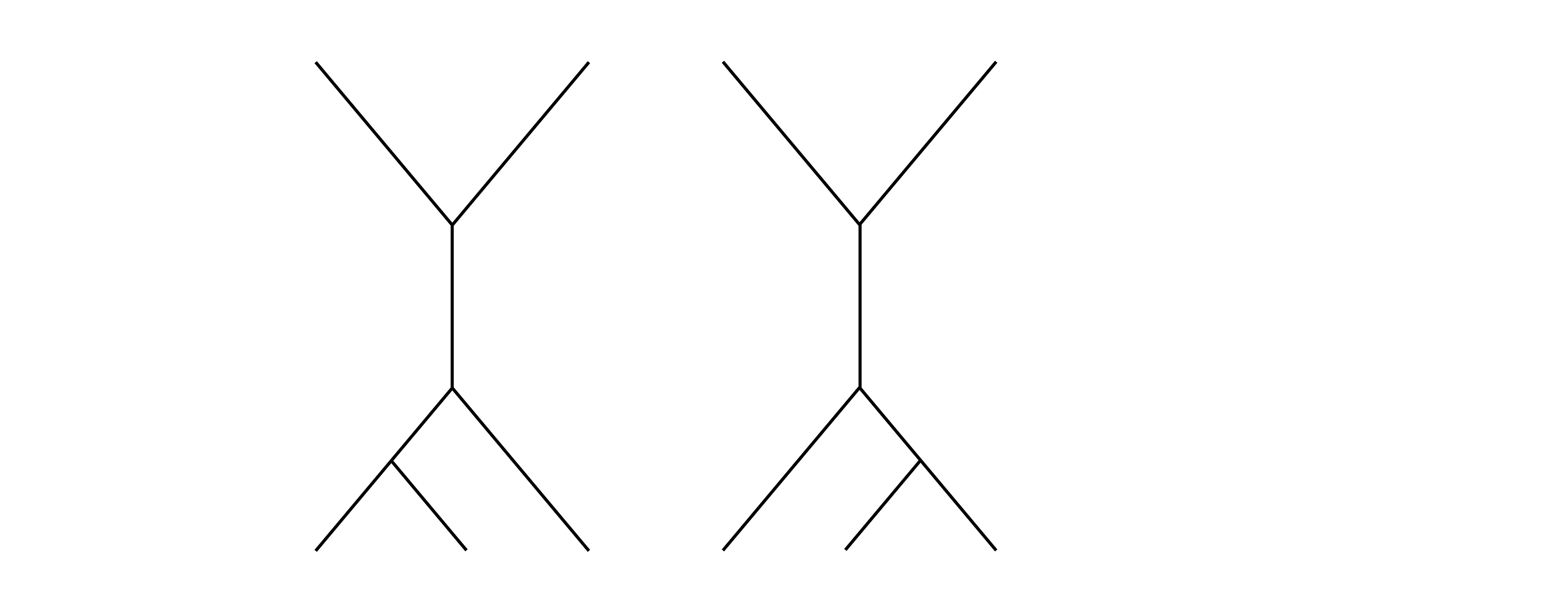}
\put(18.5,4.2){\small$1$}
\put(28.5,4.2){\small$2$}
\put(37,4.2){\small$3$}
\put(37,39.5){\small$4$}
\put(18.5,39.5){\small$5$}
\put(44.5,4.2){\small$1$}
\put(54,4.2){\small$2$}
\put(63,4.2){\small$3$}
\put(44.5,39.5){\small$5$}
\put(63,39.5){\small$4$}
\put(41,22.5){\small$+$}
\end{overpic}
\caption{Partial sum of the Feynman diagrams for  a five-point amplitude.}
\label{fig00}
\end{center}
\end{figure}

\section{Properties of urchins}
\label{sec:398jpff4}

\noindent
As we saw at the end of the previous section, 
there are  more than one ways to convert given Feynman diagrams into urchins by using \eqref{formula2} in general.
This degree of freedom leads to nontrivial relations among different 
urchins.
These relations are after all aggregated into a single identity \eqref{eq:formula3}, as we  will describe below.

From now on $ {\sum}'' $ represents the cyclic sum over the indices of external lines, and $ {\sum}$ represents the sum over the permutations of them. We also introduce $ {\sum}'$ such that
$
 {\sum}= {\sum}' {\sum}''.
$
In other words,  ${\sum}'$ represents the summation over configurations of external lines which are different modulo cyclic permutation; ${\sum}'$ represents the circular permutation. 

\subsection{Recurrence relation among urchins}
\label{sec:3.1}

\noindent
It is convenient to introduce a partial sum of Feynman diagrams $\langle x,y,z\rangle $ 
$(x,y,z \in \mathbb N)$ defined by
\begin{equation}
\langle x,y,z\rangle ={\sum}''
I\left( 
 \left(T_x\right)_{1}^{x}, 
 m\left(
  \left(T_y\right)_{x+1}^{x+y},  \left(T_z\right)_{x+y+1}^{x+y+z}
  \right)
 \right),
\end{equation}
where $(T_w)_p^{p+w-1}$ is given by
\begin{equation}
(T_w)_p^{p+w-1}=T_w(\varphi_{p},\varphi_{{p+1}},..., \varphi_{{p+w-1}}).
\end{equation}

\noindent
From the definition of $T_n(\varphi_{1},..., \varphi_{n})$, 
it follows that
\begin{equation}
\sum_{i=1}^{y-1}\langle x,i,y-i\rangle =\sum_{j=1}^{x-1}\langle y,j,x-j\rangle ,\quad x,y\ge 2.
\end{equation}

\noindent
Now define the partial sum of the urchins $ [l, m, n] $ by
\begin{equation}
{\sum} \int \left( A \mathcal O\right)_1^{l-1}
W\mathcal O_{l}
 \left( A \mathcal O\right)_{l+1}^{l+m-1}
W\mathcal O_{l+m}
 \left( A \mathcal O\right)_{l+m+1}^{l+m+n-1}
W \mathcal O_{l+m+n}
\end{equation}
where
\begin{equation}
(A\mathcal O)_{p}^{q}=A\mathcal O_pA\mathcal O_{p+1}A\mathcal O_{p+2}...A\mathcal O_q.
\end{equation}
By definition, $[x,y,z]$ is invariant under the cyclic permutation of its arguments,
\begin{equation}
\label{eq:junkai4567}
[x,y,z]=[y,z,x]. 
\end{equation}
From the formula \eqref {formula2}, $\langle x,y,z\rangle$ can be represented by using $[l, m, n]$,
\begin{equation}
\label{eq:correspondence1}
{\sum}'\langle x,y,z\rangle =[x,y,z].
\end{equation}
This leads to a relation that holds among urchins. Let $ x $, $ y $ be natural numbers equal to or greater than two. It then follows that
\begin{equation}
\label{eq:formula3}
\sum_{i=1}^{y-1}[x,i,y-i]=\sum_{j=1}^{x-1}[y, j, x-j].
\end{equation}

%
Before proceeding to the next section, 
let us define a symmetric sum of Feynman diagrams $\langle x, y\rangle $ as below 
\begin{equation}
\langle x,y\rangle ={\sum}''\sum_{i=1}^{y-1}
I\left(
  \left(T_x\right)_{1}^{x}, 
  m\left( \left(T_{y-i}\right)_{x+1}^{x+y-i}, \left(T_i\right)_{x+y-i+1}^{x+y}\right)
  \right).
\end{equation}
In other words, $ \langle x, y\rangle  $ is obtained by connecting $ \left(T_x\right)_{1}^{x} $ and $ \left(T_y \right)_{x+1}^{x+y}$ with one propagator removed:
\begin{equation}
 \langle x, y\rangle={\sum}''I(\left(T_x\right)_{1}^{x}, (\mathcal P)^{-1}\left(T_y\right)_{x+1}^{x+y}).
\end{equation}
We also define a two-headed urchin $[x,y]$ by
\begin{equation}
\label{eq:correspondence2}
[x,y]=\sum_{i=1}^{y-1}[x,y-i,i] ={\sum}'\langle x,y\rangle.
\end{equation}
By definition, it holds that
\begin{equation}
[x,y]=[y,x], 
\quad \langle x,y\rangle=\langle y,x\rangle. 
\end{equation}

Let us conclude this subsection with a remark on gauge invariance of the formula \eqref{eq:formula3}. 
We have proved the relation  \eqref{formula2} by using 
Feynman diagrams in the dressed $\mathcal B_0$ gauge, and it is assumed accordingly  that $\Psi_T$ is an Okawa-type solution and $\Psi=0$. 
Nevertheless, \eqref{eq:formula3} is valid 
 regardless of the gauge fixing conditions of $\Psi$ and $\Psi_T$. 
This can be proved by considering an infinitesimal gauge transformation of \eqref{eq:formula3} in a similar manner as in Section~4.3 of \cite{Masuda:2019rgv}.

\begin{figure}[tbp]
\begin{center}
\begin{overpic}[width=11cm, bb=0 0 1024 768
]
{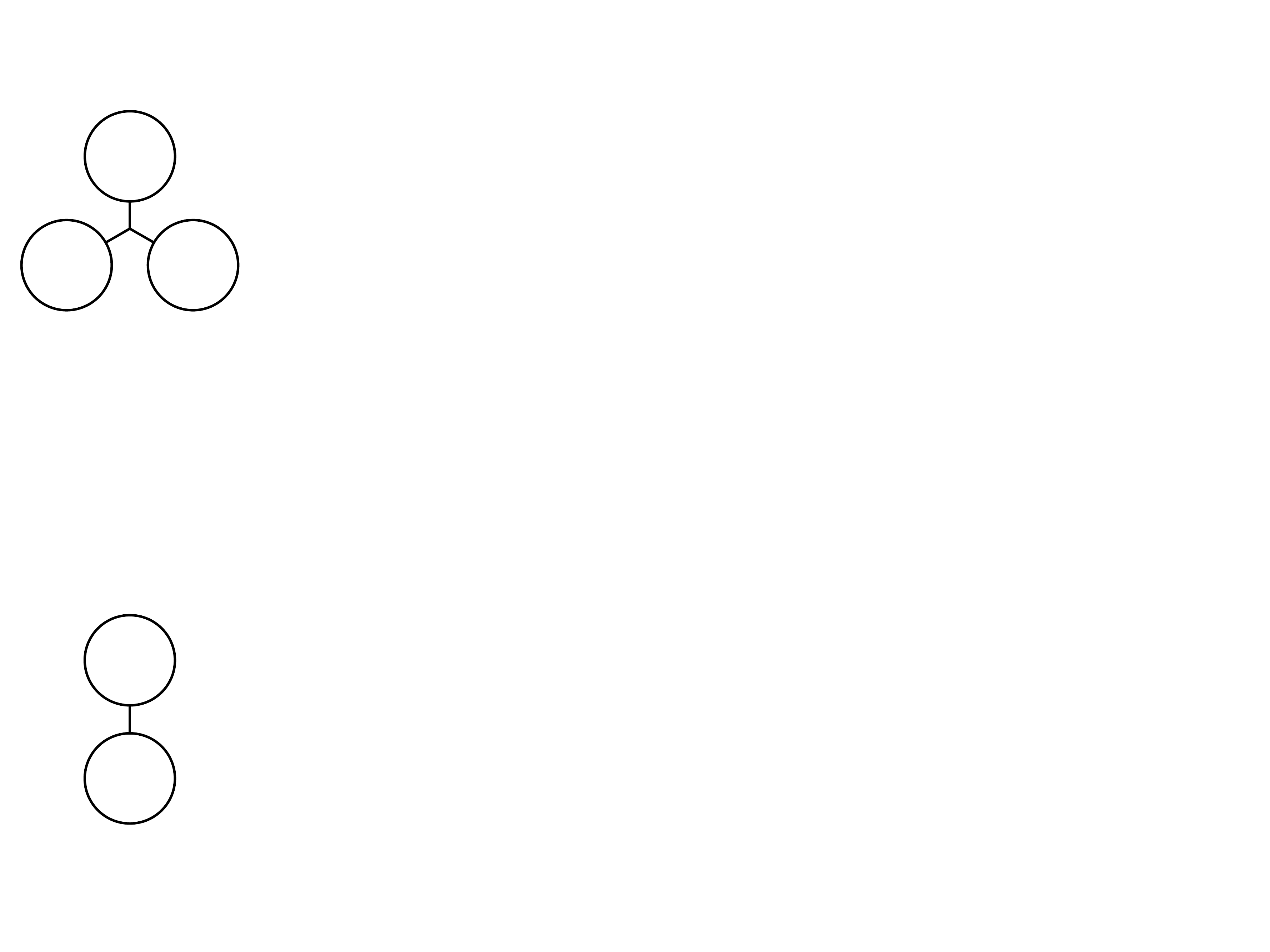}
\put(-5,70){(a)}
\put(-5,32){(b)}
\put(4.39,53.4){$x$}
\put(14.5,53.4){$y$}
\put(9.3,61.9){$z$}
\put(9.3,22.1){$x$}
\put(9.3,12.9){$y$}
\put(20,57){\small$\displaystyle =\sum\int W\mathcal O_1\left( A\mathcal O\right)_{2}^{x}
W\mathcal O_{x+1}\left( A\mathcal O\right)_{x+2}^{x+y}
W\mathcal O_{x+y+1}\left( A\mathcal O\right)_{x+y+2}^{x+y+z}
$}
\put(20,47.5){\small$\displaystyle =\sum\int \left(T_x\right)_{1}^{x} \ast \left(T_y\right)_{x+1}^{x+y}\ast \left(T_z\right)_{x+y+1}^{x+y+z}
$}
\put(20,38){\small$\displaystyle \equiv[x,y,z]
$}
\put(20,17.5){\small$\displaystyle =\sum\sum_{i=1}^{y-1}\int  \left(T_x\right)_{1}^{x}\ast \left(T_{y-i}\right)_{x+1}^{x+y-i}\ast \left(T_i\right)_{x+y-i+1}^{x+y}
$}
\put(20,9){\small$\displaystyle \equiv[x,y]
$}
\end{overpic}
\caption{(a) A schematic picture for $[x,y,z]$. The number written in each circle represents “length” of the three parts separated by $ W$ in the urchin. (b) A schematic  picture for $[x,y]$. }
\end{center}
\end{figure}

\begin{figure}[tbp]
\begin{center}
\begin{overpic}[width=11cm, bb=0 150 1024 690
]
{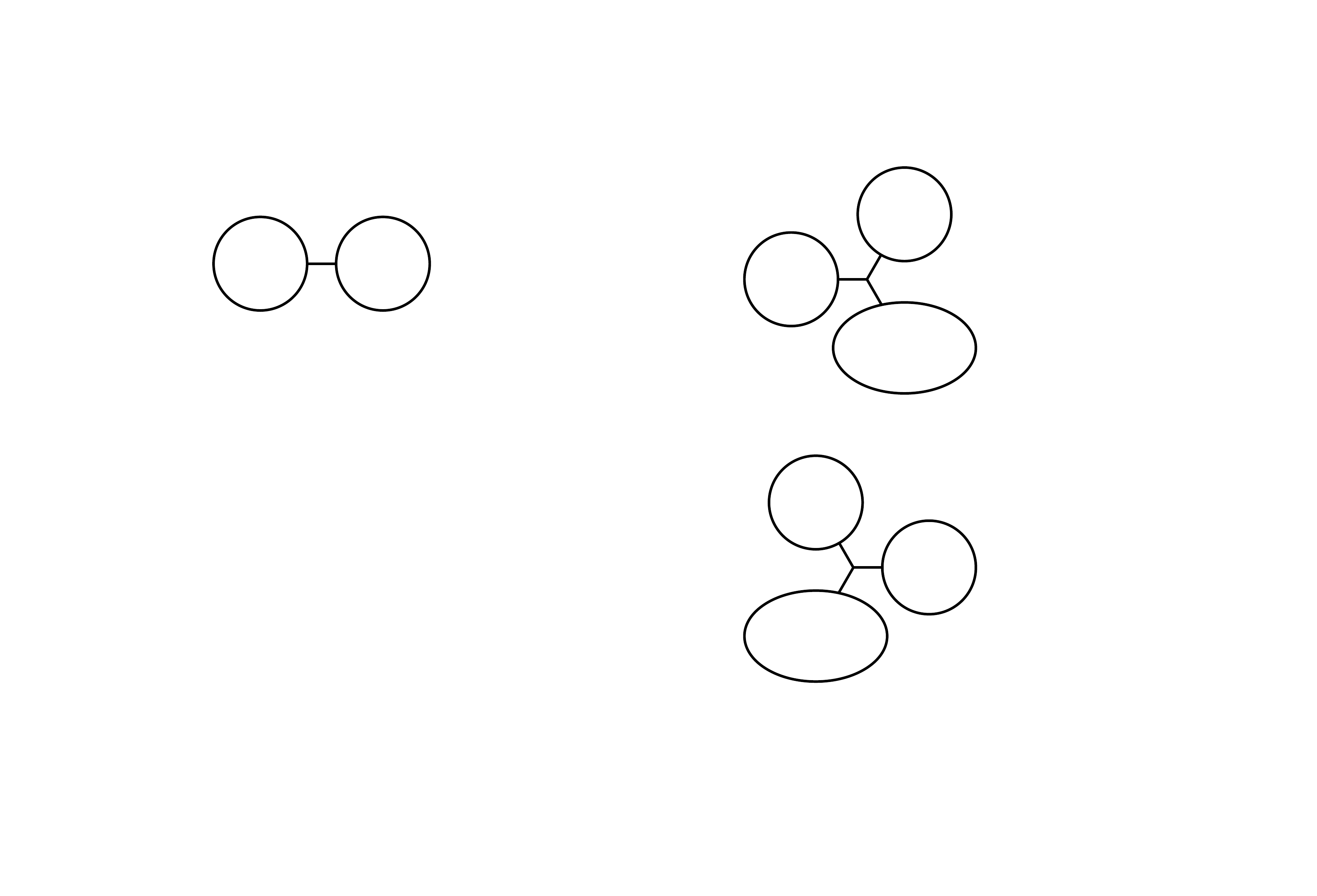}
\put(18.7,32.7){$x$}
\put(28.2,32.7){$y$}
\put(65,26.2){$y-i$}
\put(59,31.4){$x$}
\put(68,36){$i$}
\put(61,14.5){$j$}
\put(69.6,9.7){$y$}
\put(58,4.3){$x-j$}
\put(41.5,32){$\displaystyle =\sum_{i=1}^{y-1}
$}
\put(41.5,9){$\displaystyle =\sum_{j=1}^{x-1}
$}
\end{overpic}
\caption{
A schematic description for the definition of $[x,y]$
(the first equality)
and the basic relation \eqref{eq:formula3} among the three-headed urchins
(the second equality). }
\end{center}
\end{figure}

\subsection{Basis for urchins}
\label{sec:3.2}
Having obtained a relation among urchins \eqref{eq:formula3},
we now wish to find a set of linearly independent urchines. 
In the following, we do not distinguish 
$[x,y,z]$s obtained by 
a cyclic shift of their arguments; 
that is, 
 we identify $[x,y,z]$ and $[y,z,x]$.  
 Let $V_N$ be a vector space of urchins defined in this way,
\begin{equation}
V_N
=
\left\{
\sum_{x,y,z} c_{x,y,z}[x,y,z] \;\left|\;
  \begin{gathered}
  c_{x,y,z}\in \mathbb R, \quad   x,y,z\in \mathbb N,\\
    x+y+z=N, \\
    [x,y,z]\sim [y,z,x]
  \end{gathered}
\right.
\right\}.
\end{equation}
 
 \subsubsection{Dimension of $V_N$}
 \noindent
The dimension of $V_N$ is given by\footnote{This implies that growth in the number of urchins (with respect to the number of the external states $N$) is slower than growth in the number of Feynman diagrams; the number of 
the Feynman diagrams
 contained in $T_n$ is given by the Catalan number, which grows asymptotically as
\begin{equation}
C_n\sim \frac{4^n}{\sqrt{\pi}n^{\frac{3}{2}}}.
\end{equation}
} 
 \begin{equation}
 \label{eq:vavv3bduu}
 \mathrm{dim} V_N=\left\lfloor
\frac{1}{6}(N-1)(N-2)+\frac{2}{3}
\right\rfloor
-
\left\lfloor  \frac{N-3}{2}\right\rfloor. 
 \end{equation}
Here $\lfloor x\rfloor$ is the floor function, namely the maximal integer $n$ which satisfies $n\le x$. \\

\noindent
\textbf{Proof of \eqref{eq:vavv3bduu}}: 
To prove \eqref{eq:vavv3bduu}, we first count 
the number of the triplets of natural numbers $(x,y,z)$ whose total is $N$ 
under the identification $(x,y,z)\sim (y,z,x)$, 
\begin{equation}
\#(x,y,z)
:=
\#\left\{
(x,y,z) \;\left|\;
  \begin{gathered}
    x,y,z\in \mathbb N,\\
    x+y+z=N,\\
    (x,y,z)\sim (y,z,x)
  \end{gathered}
\right.
\right\}.
\end{equation}
The answer depends on whether $N$ is a multiple of three:
\begin{equation}
\label{ew:hg65rd}
\#(x,y,z)
=
\begin{cases}
\displaystyle \frac{1}{6}(N-1)(N-2)&(3\nmid N)
\vspace{3mm}\\

\displaystyle \frac{1}{6}(N-1)(N-2)+\frac{2}{3}&(3\mid N),
\end{cases}
\end{equation}
or this can be expressed  in a single line as
\begin{equation}
\#(x,y,z)
=
\left\lfloor
\frac{1}{6}(N-1)(N-2)+\frac{2}{3}
\right\rfloor. 
\end{equation}
The number of the constraints from \eqref{eq:formula3} equals 
that of $[x,y]$s with $2\le x<y$ and $x+y=N$, thus we find
\begin{equation}
\label{eq:tgh8i}
\# \textrm{(constraints from \eqref{eq:formula3})} =\left\lfloor  \frac{N-3}{2}\right\rfloor. 
\end{equation}
The difference between \eqref{ew:hg65rd} and \eqref{eq:tgh8i}  therefore gives the number of linearly independent urchins. $\square$

\subsubsection{A basis for $V_N$}
\noindent
The following form a basis for $V_N$:
\begin{equation}
\label{eq:basis1}
\big\{ \, [x,y,z] \, \big|\,  2\le x,y,z,\ x+y+z=N  \,\big\}\cup \big\{\, [1,1,N-2] \, \big\}\cup \big\{\, v^-_{i,j}\, \big\},
\end{equation}
where
\begin{equation}
v^-_{i,j}=[1,i,j]-[1,j,i]\qquad (2\le i< j),
\end{equation}
\begin{equation}
v^+_{i,j}=[1,i,j]+[1,j,i]\qquad (2\le i\le j).
\end{equation}
Indeed, the number of the elements in \eqref{eq:basis1} corresponds to \eqref{eq:vavv3bduu}.  Also, we can always erase $v^+_{i,j}$ by using the following formula:
%
\begin{equation}
\label{eq:ueno37hf89}
v_{i,N-i-1}^+=\frac{1}{3}\sum_{\substack{2\le p, q, r\le N-4\\p+q+r=N}}\left(\rho_i(p)+\rho_i(q)+\rho_i(r)\right)[p,q,r]+[1,1,N-2]
\end{equation}
where 
\begin{equation}
\rho_i(x)=\Theta(N-i \le x\le N-4)-\Theta(2\le x\le i)
\end{equation}
with $\Theta(p)$ a Boolean-valued function
\begin{equation}
\Theta(\text{statement})=
\begin{cases}
1&\text{if statement is true;}\\
0&\text{if statement is false.}
\end{cases}
\end{equation}
\textbf{Proof of \eqref{eq:ueno37hf89}} :   Let us rewrite \eqref{eq:formula3} as 
\begin{align}
v_{2,N-3}^++K(2)&=[1,1,N-2]\notag\\
v_{3,N-4}^++K(3)&=v_{2,N-3}^+\notag\\
v_{4,N-5}^++K(4)&=v_{3,N-4}^++K(N-4)\notag\\
&\vdots \notag\\
\label{eq:cdf4ufxc543}
v^+_{i,N-i-1}+K(i)&=v^+_{N-i,i-1}+K(N-i)
\end{align} 
where $K(i)$ is defined by
\begin{equation}
\label{eq:K(i)}
K(i)=
\begin{cases}
\displaystyle\sum_{k=2}^{N-i-2}[i,k,N-i-k]\quad &(1\le i\le N-4)\\
0\quad &\text{(otherwise)}.
\end{cases}
\end{equation}
Another useful expression for $K(i)$ is
\begin{equation}
\label{eq:u478cgbe}
K(l)=\frac{1}{3}\sum_{\substack{2\le p,q,r\le N-4\\p+q+r=N}}(\delta_{l,p}+\delta_{l,q}+\delta_{l,r})[p,q,r]. 
\end{equation}
By summing the equations in \eqref{eq:cdf4ufxc543}, we obtain that\footnote
{We follow the convention that 
when the upper limit of the summation index is lower than the lower limit,  the sum is zero
\begin{equation}
\sum_{i=a}^b(...)=0\quad (a>b). 
\end{equation}
}
\begin{equation}
v_{i,N-i-1}^+=\sum_{l=N-i}^{N-4} K(l)-\sum_{k=2}^i K(k)+[1,1,N-2],
\end{equation}
and by using the expression \eqref{eq:u478cgbe} we obtain \eqref{eq:ueno37hf89}.
$\square$\\

\subsection{Proof 
that $I_\Psi^{(N)}$ is the $S$-matrix}
\label{sec:section3.3}

\noindent
Now, we shall prove that $I_\Psi^{(N)}$ in \eqref{eq:fml} represents the on-shell scattering amplitude.
On one hand, $S$-matrix  in the dressed $\mathcal B_0$ gauge in the tree level is given by
\begin{equation}
\begin{split}
\label{eq:1111111111}
{\mathcal A}^{(N)}(\varphi_1,...,\varphi_N)
&={\sum}' I(\varphi_N,{\mathcal P}^{-1}T_{N-1}(\varphi_1,... , \varphi_{N-1}))\\
&=-\frac{1}{N}[1,N-1].
\end{split}
\end{equation}
The factor ${1}/{N}$ is due to an extra summation over cyclic permutation of the external lines in the definition of $[1,N-1]$.  
We expand this quantity
in terms of 
the basis \eqref{eq:basis1}:
\begin{equation}
\label{eq:btt1rfy9}
\begin{split}
[1,N-1]=&[1,1,N-2]+[1,2,N-3]+...+[1,N-3,2]+[1,N-2,1]\\
=&2[1,1,N-2]+\frac{1}{2}\sum_{i=2}^{N-3}v_{i,N-i-1}^+\\
=&\frac{N}{2}[1,1,N-2]
-\frac{N}{6}\sum_{\substack{p+q+r=N\\2\le p,q,r\le N-4}}  [p,q,r].
\end{split}
\end{equation}
Here we used \eqref{eq:ueno37hf89} in the last equality. 

On the other hand, from \eqref{eq:fml}, $I_0^{(N)}$ is given by
\begin{equation}
\label{eq:2222222}
I_0^{(N)}=\frac{1}{3N}C_N\sum_{\substack{p+q+r=N\\1\le p,q,r\le N-2}} [p,q,r],
\end{equation}
which can be written as
\begin{equation}
\begin{split}
=&\frac{1}{3N}C_N\left(3[1,1,N-2]+\frac{3}{2}\sum_{i=2}^{N-2}v_{i,N-i-1}^++\sum_{\substack{p+q+r=N\\2\le p,q,r\le N-4}}[p,q,r] \right)\\
=&\frac{1}{N}C_N\Bigg(\frac{N-2}{2}[1,1,N-2]-\frac{N-2}{6}\sum_{\substack{p+q+r=N\\2\le p,q,r\le N-4}}  [p,q,r]\Bigg)
\end{split}
\end{equation}
where we again used \eqref{eq:ueno37hf89} in the last equality. 
Therefore, we conclude that $I_0^{(N)}$ is the $S$-matrix  if we chose $C_N$ to be
\begin{equation}
C_N=-\frac{N}{N-2}.
\end{equation}

%
%
%
%
%
%
%
%
%
%


\section{$S$-matrix from the new set of Feynman rules presented in \cite{Masuda:2019_2}
}
\label{sec:4}

In this section, we will review the new set of Feynman rules $\mathcal R_\flat$
 introduced in \cite{Masuda:2019_2}, and then prove that the result of $\mathcal R_\flat$ matches $I^{(N)}_\Psi$. 
Throughout this section, we will choose $A$ as
\begin{equation}
\label{eq:8sru9087edg}
A=\frac{W_\Psi A_T}{1+W_\Psi}
\end{equation}
 with $A_T$ satisfying $A_T^2=0$, following \cite{Masuda:2019_2}.\footnote{
 In the context of \cite{Masuda:2019_2}, the right hand side of \eqref{eq:8sru9087edg} is interpreted in terms of formal power series.  
In the present context, however, we can define the right hand side of \eqref{eq:8sru9087edg} by using a superposition of wedge states, and 
Feynman diagrams of $\mathcal R_\flat$ are calculable in general.   
Note that the inverse of $K$ appearing there is treated by the method of \cite{Masuda:2019rgv}.
} 
Under this choice, $A$ and $W_\Psi$ satisfies 
$A^2=0$ and $[A, W_\Psi]=0$.

\subsection{Summary of the new Feynman rules $\mathcal R_\flat$}

\noindent
Inspired by the formula $I_\Psi^{(N)}$, a  new propagator ${\mathcal P}_\flat$ is devised 
in \cite{Masuda:2019_2}, 
which reads\footnote
{For general cases, where $A$ does not commute with $W_\Psi$, we can take ${\mathcal P}_\flat$ as
\begin{equation}
{\mathcal P}_\flat \phi=\frac{1}{2}\left( \frac{1}{\sqrt{W_\Psi}}{A}\frac{1}{\sqrt{W_\Psi}} \ast \phi+(-1)^{\text{gh}(\phi)} \phi\ast \frac{1}{\sqrt{W_\Psi}}{A}\frac{1}{\sqrt{W_\Psi}}\right), 
\end{equation}
and the discussion in this section still holds. }
\begin{equation}
\label{eq:64ryfh}
{\mathcal P}_\flat\phi=\frac{A}{2W_\Psi}\ast \phi+(-1)^{\text{gh}(\phi)} \phi\ast \frac{A}{2W_\Psi}. 
\end{equation}
Here $\phi$  denotes a test state and $\text{gh}(\phi)$ the world-sheet ghost number of $\phi$. 
%

Validity of ${\mathcal P}_\flat$ is discussed in the paper \cite{Masuda:2019_2} by using the homological perturbation.  
But the authors did not prove validity of some assumptions there, including that on the physical states. 
In particular, when we choose the Okawa-type solution, validity of their argument is not clear due to the inverse of $K$ (see Appendix \ref{sec:c2}), which is hidden 
behind a formal power series that may not converge. 
The validity of the perturbation theory using ${\mathcal P}_\flat$ is therefore not yet 
established. 
%
Also, the gauge fixing condition or the physical states for $\mathcal R_\flat$ have not been clarified. 

In the following discussion, which is independent of that of \cite{Masuda:2019_2}, we only assume that the external state $\varphi_j$ and 
$\mathcal O_j$ are related by $\varphi_j=\sqrt{-W_\Psi}\mathcal O_j\sqrt{-W_\Psi}$, and 
do not assume anything about $\mathcal O_j$ except that $\mathcal O_j$ is a $Q_\Psi$-closed state at ghost number 1. 
We do not need to specify the gauge fixing condition of $\varphi_j$ in the following discussion.\\

\noindent
\textbf{Example:} Let us calculate an on-shell five-point amplitude with $\mathcal R_\flat$. Since all the five Feynman diagrams are topologically equivalent, the five-point amplitude can be written as
\begin{equation}
\notag
\begin{split}
\sum I(m(\varphi_4,\varphi_5), {\mathcal P}_\flat m({\mathcal P}_\flat m(\varphi_1,\varphi_2), \varphi_3))
=-\frac{1}{4}\left(2[2,2,1]+[3,1,1]\right). 
\end{split}
\end{equation}
We can confirm that this expression is equivalent to \eqref{eq:fml} by using \eqref{eq:formula3}. \\

\noindent
In the next subsection, we will evaluate the Feynman diagrams in $\mathcal R_\flat$ and show that the sum of them equals the on-shell scattering amplitude. 
\footnote{
It can generally be proved that the on-shell scattering amplitude does not depend on 
the choice of the propagator. 
However, 
it is quite unclear whether this argument holds in the presence of $B/K$. 
This is why we will evaluate the Feynman diagrams explicitly in Section \ref{sec:ddggdfdlllp3}. 
We will discuss this point in more detail in Appendix \ref{sec:c2}. 
}


\subsection{Proof 
that $\mathcal R_\flat$ gives the correct tree-level amplitudes
}
\label{sec:ddggdfdlllp3}
We divide the proof into three steps. First, we classify the non-vanishing Feynman diagrams. Then, we evaluate the sum of each class of the Feynman diagrams. Finally, we prove that the sum of the Feynman diagrams equals~\eqref{eq:fml}.

\subsubsection{A classification of non-vanishing diagrams}
\noindent
A notable feature of $\mathcal R_\flat$ is that  the Feynman diagrams satisfying a  specific condition always vanish. 
Let us describe this condition precisely. 
Let $p(g)$ be a subdiagram of a Feynman diagram $g$ 
which is obtained by getting rid of all the
external lines of $g$.
If the number of the branching points in $p(g)$ is greater than one, 
then the contribution of $g$ is zero. 
This is because $g$ has more than three $W_\Psi$s when it is expressed by using $W_\Psi$, $A$, and $\mathcal O_j$, which implies collision of $A$s.\footnote{
In the case where we chose $A$ which does not satisfy $A^2=0$, let us say $A= \frac{W_\Psi A_T}{W_\Psi+1}+Q\chi$,  
we can still prove that the Feynman diagrams satisfying the above condition vanish as a result of partial integration with respect to $Q$.  }

We therefore only need to consider the Feynman diagrams of which $p(g)$ does not have branching point (A-type) or those of which $p(g)$ has one branching point (B-type). 

\subsubsection{Evaluation of the Feynman diagrams}

\noindent
We first present the formula for the sum of the Feynman diagrams of each type, and then move on to the proof.  
The contribution from the Feynman diagrams of A-type is
\begin{equation}
\label{eq:a-type}
-
\left(\frac{1}{2}\right)^{N-3}\sum_{p=0}^{N-4} 
\begin{pmatrix}
N-4\\
p
\end{pmatrix}
[N-p-2,p+2]
\end{equation}
while the contribution from the Feynman diagrams of B-type is
\begin{equation}
\label{eq:b-type}
-
\frac{1}{3}\left(\frac{1}{2}\right)^{N-3}\sum_{\substack{p+q+r=N-6\\0\le p,q,r}} 
2f(p,q,r)
[p+2,q+2,r+2]
\end{equation}
where
\begin{equation}
f(p,q,r)=\sum_{p_1=0}^p\sum_{q_1=0}^q\sum_{r_1=0}^r
\begin{pmatrix}
p_1+r-r_1\\
p_1
\end{pmatrix}
\begin{pmatrix}
q_1+p-p_1\\
q_1
\end{pmatrix}
\begin{pmatrix}
r_1+q-q_1\\
r_1
\end{pmatrix}.
\end{equation}

\noindent
\textbf{Proof of \eqref{eq:a-type} and \eqref{eq:b-type}:} To calculate the sum of all the A-type Feynman diagrams, it is convenient 
to consider the following three phases in constructing and evaluating Feynman diagrams:
\\

\noindent
\textbf{Phase 1.}  We first set the numbers of external lines on each side of $p(g)$. See Figure 6. 
The red segment represents $p(g)$. Let $p$ be the number of external lines on one side of $p(g)$, which are represented by purple lines. Accordingly, there are $N-p-4$ external lines on the other side, which are represented by pale-blue lines. 
\\

\noindent
\textbf{Phase 2.} We then attach these external lines to $p(g)$. See Figure 8. There are 
 \begin{equation}
 \label{eq:binomi1}
 \begin{pmatrix}
 N-4\\
 p
 \end{pmatrix}
 \end{equation}
ways to do this. 
\\

\noindent
\textbf{Phase 3.} Now, we convert a Feynman diagram into the sum of urchins. 
  For every propagator ${\mathcal P}_\flat$, we draw a dotted yellow line, which is rooted in ${\mathcal P}_\flat$ and growing on either side of it. See Figure \ref{fig:hhh}. 
This dotted yellow line represents a factor of $\frac{1}{2}A$. 
As exemplified in Figure \ref{fig:hhh3}, 
there comes a dotted lines between every neighbouring pair of external lines, except for only three pairs.  
(To obtain a  non-vanishing Feynman diagram, two $A$s must not collide.)
We put $W_\Psi$s in these three empty places. 
The two of these three $W_\Psi$s are always placed at both ends of $p(g)$. 
From this figure, we can read off the corresponding expression as an urchin; we only need to  place $\mathcal O_j$s, $A$s, and $W_\Psi$s in the same order as in the figure and put the integration symbol $\int$ with the factor $\left(\frac{1}{2}\right)^{N-3}$. 
Note that, if the position of three $W_\Psi$s is given, the configuration of the dotted lines is uniquely determined. 
Therefore, summing all possible configurations of dotted lines yields the following result:
\begin{equation}
\begin{split}
&-\left(\frac{1}{2}\right)^{N-3}\left(\sum_{i=1}^{N-p-3} [p+2,N-p-2-i,i]+\sum_{j=1}^{p+1} [j,p+2-j,N-p-2]\right)\\
=&-\left(\frac{1}{2}\right)^{N-3} 2[p+2,N-p-2].
\end{split}
\end{equation}

\noindent
Finally,  by taking the factor \eqref{eq:binomi1} into account and summing over $p$, we obtain the formula \eqref{eq:b-type}. \\

\noindent
Next, let us calculate the sum of all the B-type Feynman diagrams in a similar method. \\

\noindent
\textbf{Phase 1b. }
Let us first set the number of the external lines which are on each side of every branch of $p(g)$. We express these numbers by using the six variables $(p,q,r; p_1,q_1,r_1)$ as in Figure \ref{fig:fpqr}.
\\

\noindent
\textbf{Phase 2b.} We then attach the external lines to $p(g)$. There are 
 \begin{equation}
 \label{eq:binomi2}
\begin{pmatrix}
p_1+r-r_1\\
p_1
\end{pmatrix}
\begin{pmatrix}
q_1+p-p_1\\
q_1
\end{pmatrix}
\begin{pmatrix}
r_1+q-q_1\\
r_1
\end{pmatrix}
 \end{equation}
ways to do this. 
\\

\noindent
\textbf{Phase 3b.} Now, we convert a Feynman diagram into the sum of urchins. 
 We draw a dotted yellow line for every propagator ${\mathcal P}_\flat$ as before.  
For given $(p,q,r;p_1,q_1,r_1)$, there are exactly two possible configurations of the dotted lines. 
Therefore, summing these two configurations of yellow dotted lines yields the following result:
\begin{equation}
\begin{split}
&\left(\frac{1}{2}\right)^{N-3}2 [p+2,q+2,r+2].
\end{split}
\end{equation}

\noindent
Finally,  by taking the factor \eqref{eq:binomi2} into account  and summing over $(p,q,r;p_1,q_1,r_1)$, we obtain the formula (82). $\square$

\begin{figure}[p]
\begin{center}
\begin{overpic}[width=11cm, 
bb=0 0 1024 360, 
]{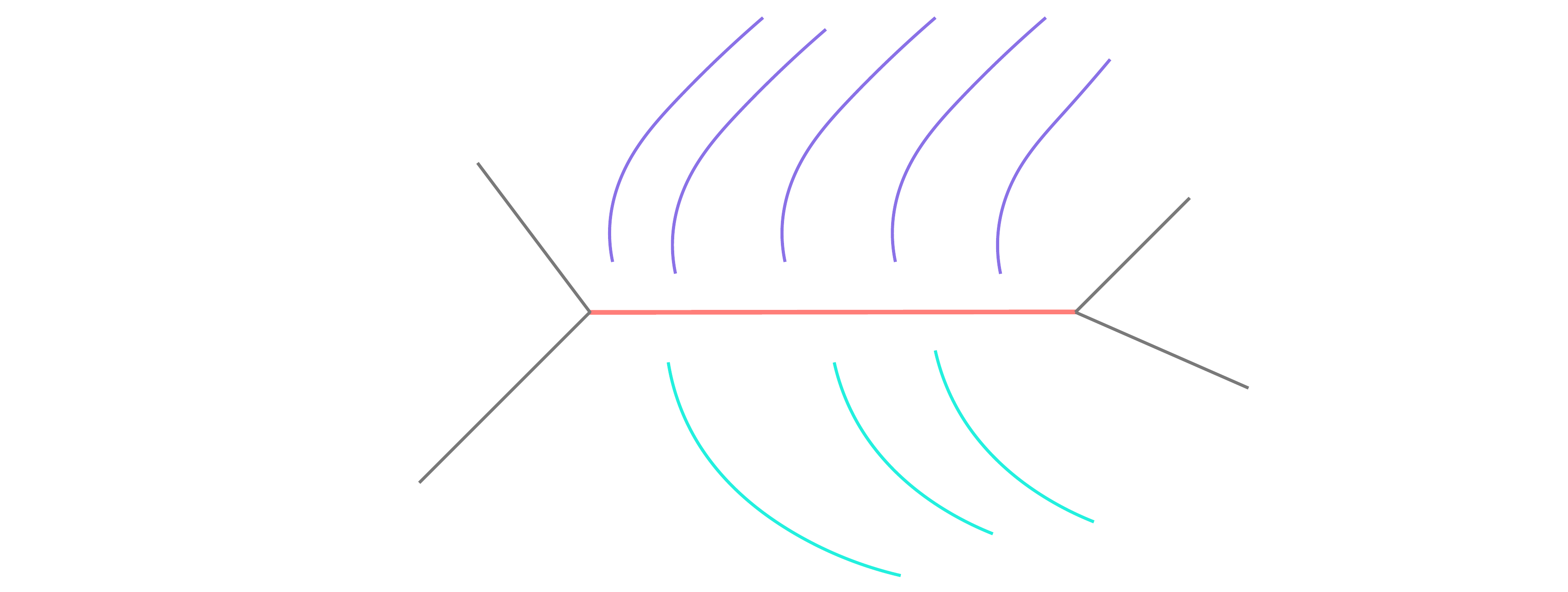}
\end{overpic}
\caption{A schematic picture for counting procedure of the A-type Feynman diagrams, phase 1. The red segment represents $p(g)$. The black, purple, and pale-blue lines represent the external states, of which there are $N$ in total. }
\end{center}
\end{figure}

\begin{figure}[p]
\begin{center}
\begin{overpic}[width=11cm, 
bb=0 0 1024 360, 
]{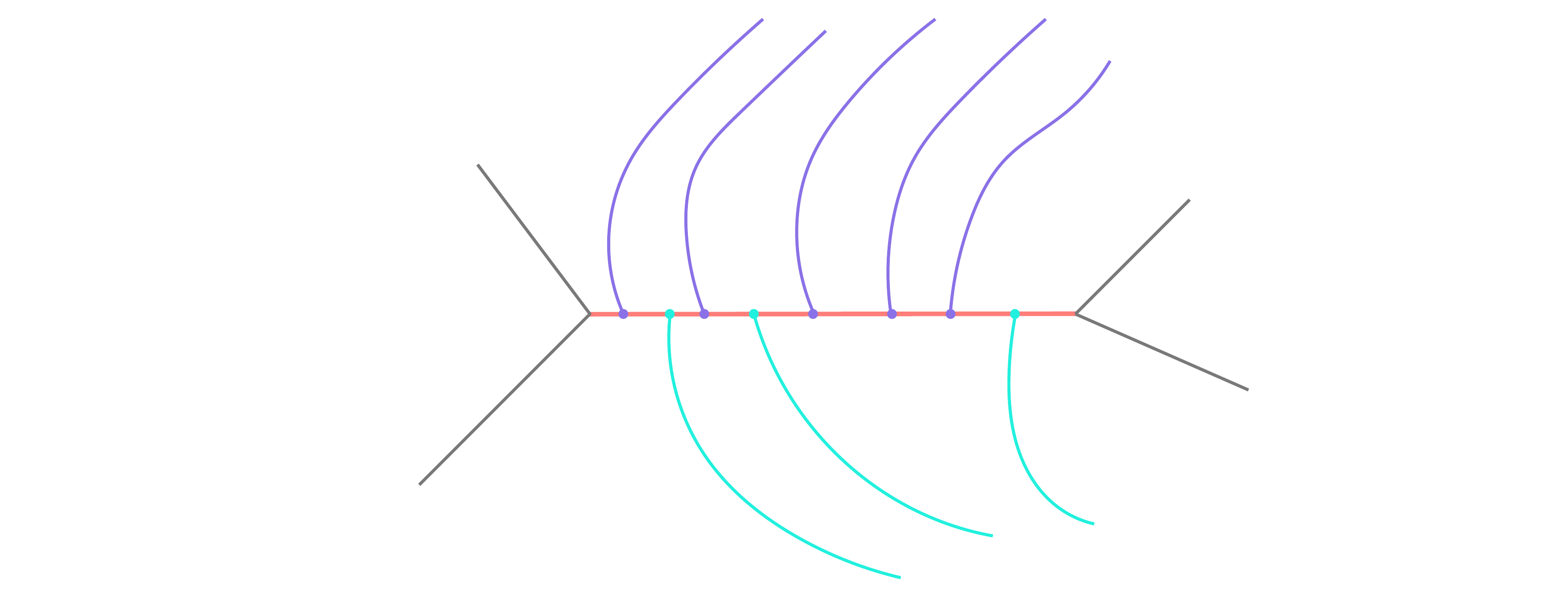}
\end{overpic}
\caption{Counting procedure of the A-type Feynman diagrams, phase 2. The external lines are connected to $p(g)$. }
\end{center}
\end{figure}

\begin{figure}[p]
\begin{center}
\begin{overpic}[width=11cm, 
bb=0 0 1024 340, 
]{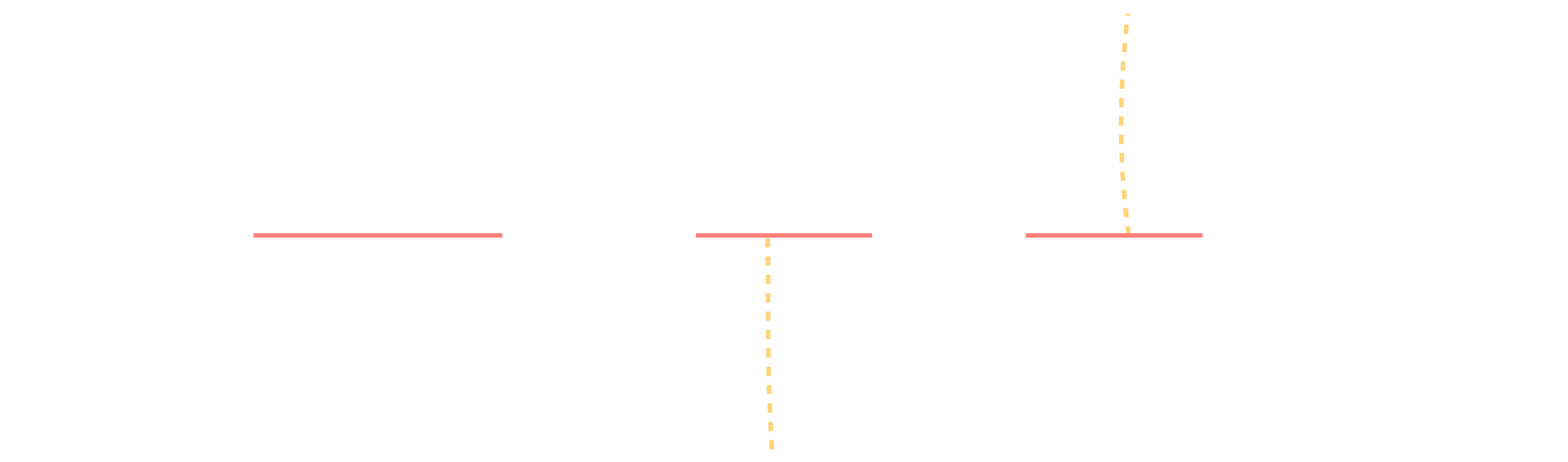}
\put(36.4,18){\small$\to$}
\put(59,18){\small or}
\put(47.8,7){\small  $A$}
\put(70.5,27){\small  $A$}
\end{overpic}
\caption{The propagator ${\mathcal P}_\flat$ and $A$. The doted line represents the position where  $A$ is inserted when the Feynman diagram is expressed as an urchin. }
\label{fig:hhh}
\end{center}
\end{figure}

\begin{figure}[tbp]
\begin{center}
\begin{overpic}[width=11cm, bb=0 200 1024 600, 
]{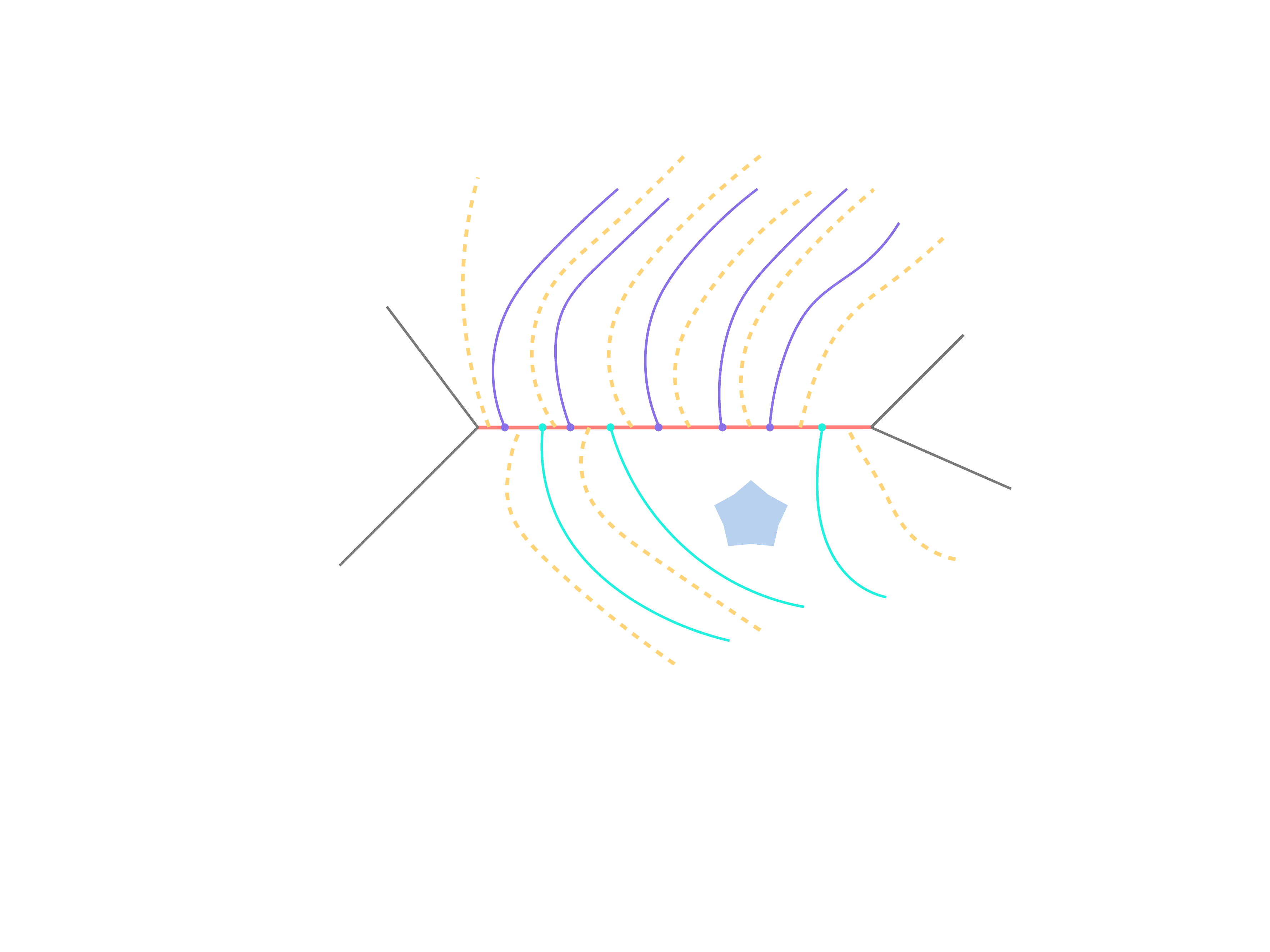}
\put(57.8,14){\scriptsize $W_\Psi$}
\put(31,22){\scriptsize $W_\Psi$}
\put(73,22){\scriptsize $W_\Psi$}
\put(38,17){\scriptsize $A$}
\put(45,17){\scriptsize $A$}
\put(68,17){\scriptsize $A$}
\put(36,30){\scriptsize $A$}
\put(41,30){\scriptsize $A$}
\put(47,30){\scriptsize $A$}
\put(54,30){\scriptsize $A$}
\put(60,30){\scriptsize $A$}
\put(66,30){\scriptsize $A$}
\end{overpic}
\caption{Counting procedure of the A-type Feynman diagrams, phase 3. 
}
\label{fig:hhh3}
\end{center}
\end{figure}

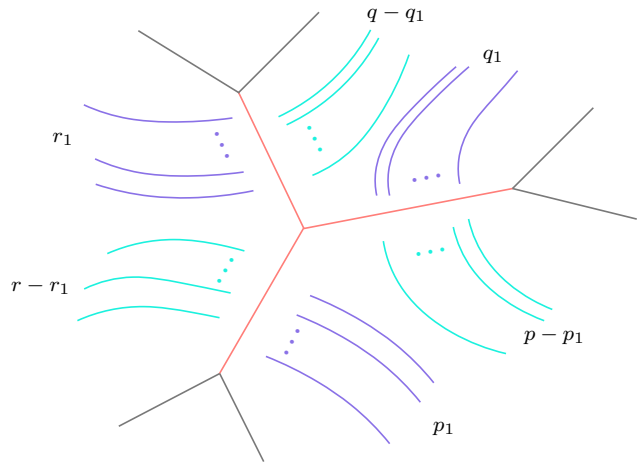
\begin{figure}[tbp]
\begin{center}
\begin{overpic}[width=11cm, bb=0 10 1024 668,
]{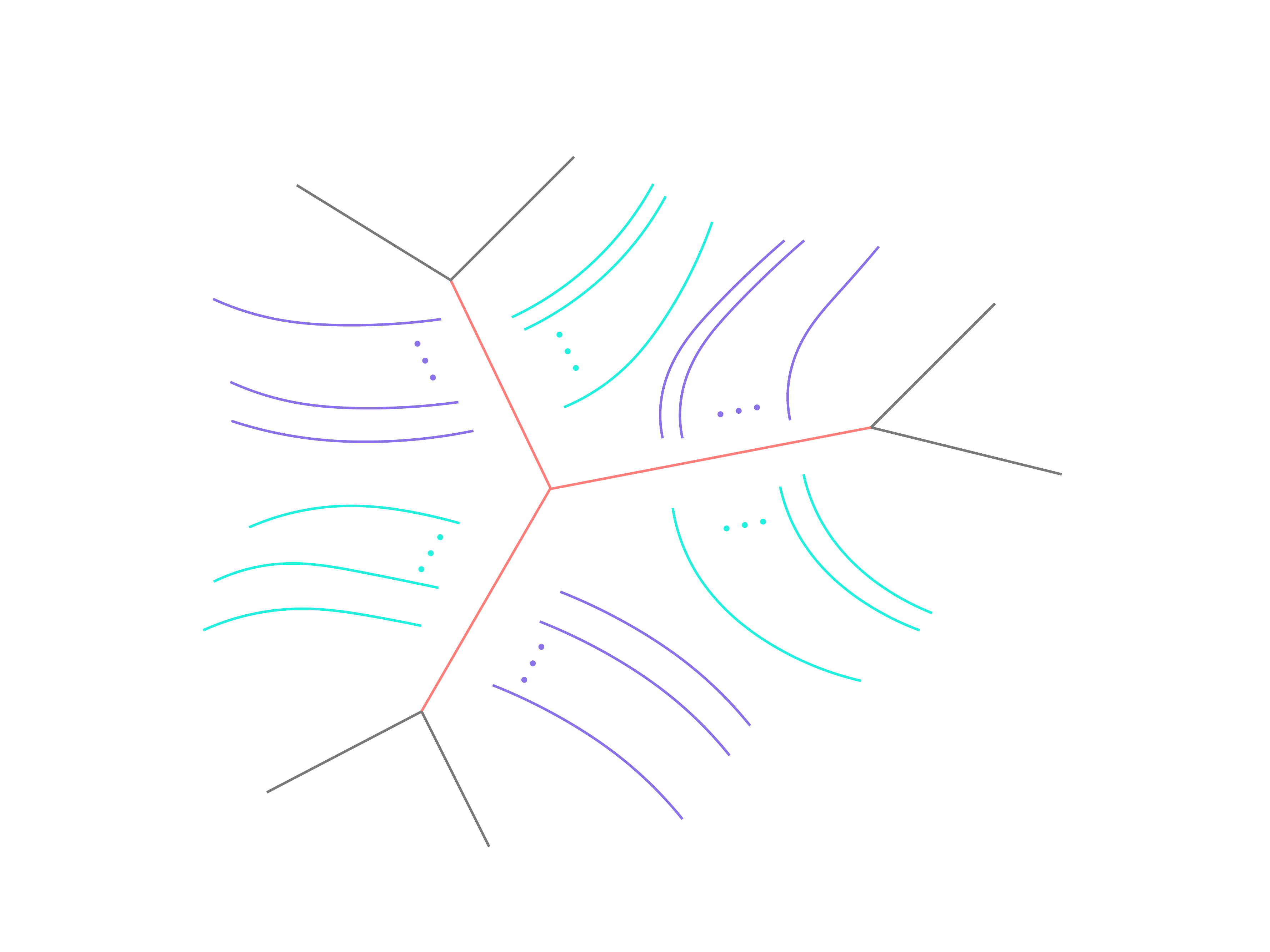}
\put(59,11){\scriptsize $p_1$}
\put(65,56){\scriptsize $q_1$}
\put(13,46){\scriptsize $r_1$}
\put(8,28){\scriptsize $r-r_1$}
\put(70,22){\scriptsize $p-p_1$}
\put(51,61){\scriptsize $q-q_1$}
\end{overpic}
\caption{A schematic picture for counting procedure of the B-type Feynman diagrams with $f(p,q,r)$. The red graph represents $p(g)$.  }
\label{fig:fpqr}
\end{center}
\end{figure}


\subsubsection{Proof of the equality}

\noindent
We are left to prove that 
the sum of \eqref{eq:a-type} and \eqref{eq:b-type} equals 
\begin{equation}
\label{eq:goi}
I_\Psi^{(N)}=-\frac{1}{2(N-3)}\sum_{i=2}^{N-2} [i,N-i].
\end{equation}
To prove this, we expand \eqref{eq:a-type}, \eqref{eq:b-type}, and \eqref{eq:goi} by using the basis of urchines \eqref{eq:basis1}. 
\noindent
The urchine $[i,N-i]$ $(2\le i\le N-2)$ is expanded in this basis as
\begin{equation}
\label{eq:orororor}
[i,N-i]=[1,1,N-2]+\frac{1}{3}\sum_{\substack{p+q+r=N\\2\le p, q, r\le N-4}}\left(\nu_i(p)+\nu_i(q)+\nu_i(r)\right)[p,q,r],
\end{equation}
where 
\begin{equation}
\nu_i(x)=\Theta(N-i \le x\le N-3)-\Theta(2\le x\le i-1).
\end{equation}
\ \\
\textbf{Proof of \eqref{eq:orororor}} :  
From the basic relation \eqref{eq:formula3} we have
\begin{equation}
[i+1,N-i-1]-[i,N-i]=K(i)-K(N-i-1),
\end{equation}
for $ 2\le i \le N-3$, and therefore
\begin{equation}
[i,N-i]=[2,N-2]+\sum_{j=N-i}^{N-3}K(j)-\sum_{k=2}^{i-1}K(k).
\end{equation}
By using \eqref{eq:u478cgbe} we obtain \eqref{eq:orororor}. $\square$
\\

\noindent
From \eqref{eq:orororor}, we find that\footnote{Here we have used
\begin{equation}
\sum_{x=0}^{N-4}\binom{N-4}{x}\nu_{x+2}(p)=2\sum_{x=0}^{p-2}\binom{N-4}{x}-2^{N-4}.
\end{equation}
}
\begin{equation}
\label{eq:vbvuyrs}
\begin{split}
&\eqref{eq:goi}-\eqref{eq:a-type}\\
=&
\frac{1}{3}\left(\frac{1}{2}\right)^{N-4}\sum_{\substack{p+q+r=N\\2\le p, q, r\le N-4}} 
\left(\sum_{x=0}^{p-2}+\sum_{x=0}^{q-2}+\sum_{x=0}^{r-2}\right)\binom{N-4}{x}[p,q,r]\\
&-\frac{1}{3}\sum_{\substack{p+q+r=N\\2\le p, q, r\le N-4}} 
[p,q,r].
\end{split}
\end{equation}
%
With the help of an identity for the binomial coefficients
\footnote{
This identity follows from the Chu-Vandermonde identity. Or, the following identities would be more fundamental: \begin{equation}
\notag
\sum_{p_1=0}^{p+q+r}\sum_{q_1=0}^{p+q+r}\sum_{r_1=0}^{p+q+r}
\begin{pmatrix}
p_1+r-r_1\\
p_1
\end{pmatrix}
\begin{pmatrix}
q_1+p-p_1\\
q_1
\end{pmatrix}
\begin{pmatrix}
r_1+q-q_1\\
r_1
\end{pmatrix}
=2^{p+q+r+2}-\delta_{p,0}-\delta_{q,0}-\delta_{r,0},
\end{equation}
\begin{equation}
\notag
\sum_{p_1=p+1}^{p+q+r}\sum_{q_1=q+1}^{p+q+r}\sum_{r_1=r+1}^{p+q+r}
\begin{pmatrix}
p_1+r-r_1\\
p_1
\end{pmatrix}
\begin{pmatrix}
q_1+p-p_1\\
q_1
\end{pmatrix}
\begin{pmatrix}
r_1+q-q_1\\
r_1
\end{pmatrix}
=0,
\end{equation}
\begin{equation}
\notag
\sum_{p_1=0}^{p}\sum_{q_1=q+1}^{p+q+r}\sum_{r_1=0}^{p+q+r}
\begin{pmatrix}
p_1+r-r_1\\
p_1
\end{pmatrix}
\begin{pmatrix}
q_1+p-p_1\\
q_1
\end{pmatrix}
\begin{pmatrix}
r_1+q-q_1\\
r_1
\end{pmatrix}
=
\sum_{j=0}^p
\begin{pmatrix}
p+q+r+2\\
j
\end{pmatrix}
-
\delta_{r,0}.
\end{equation}
The original identity is obtained by adding these identities by shifting the summation indices. 
}
\begin{equation}
f(x,y,z)+\sum_{j=0}^{x}\begin{pmatrix}n\\j\end{pmatrix}+\sum_{j=0}^{y}\begin{pmatrix}n\\j\end{pmatrix}+\sum_{j=0}^{z}\begin{pmatrix}n\\j\end{pmatrix}=2^{n}
\end{equation}
where 
$n=x+y+z+2$, 
we find \eqref{eq:vbvuyrs} equals \eqref{eq:b-type},
which completes our proof.

\section{Concluding remarks}
\label{sec:conclusion}
\noindent
We have proved that the gauge invariant quantity $ I_0^{(N)} $ equals  the sum of the tree-level Feynman diagrams in the dressed $ \mathcal B_0 $ gauge. 
By combining our results with those of \cite{Kiermaier:2007jg}, 
we conclude that $ I_0^{(N)}$ correctly reproduces the tree-level $S$-matrix. 

We have only considered the gauge invariant quantity $ I_0^{(N)} $ around the perturbative vacuum $\Psi=0$ in Sections 2 and 3. This restriction is for sake of simplicity;  
indeed, our proof can be extended to general classical solutions $\Psi$ in a straightforward manner. 
As showed in Section 5 of \cite{Masuda:2019rgv}, 
 we can calculate  $ I_{\Psi_\text{EM}}^{(N)} $ around the Erler-Maccaferri solution $\Psi_\text{EM}$ \cite{Erler:2014eqa, Erler:2019fye} essentially in the same way as  $ I_{0}^{(N)} $ . 
In particular, the relation \eqref{formula2} is appropriately generalized, and the discussion in Sections 3 is also valid in this case. Assuming every clasiccal solution is gauge equivalent to an Erler-Maccaferri solution, we can conclude that $I_\Psi^{(N)}$ is the tree-level $S$-matrix. 

The dressed  $\mathcal B_0$ gauge is regarded as a singular limit of a series of regular gauge fixing conditions, as discussed for the Schnabl gauge in  \cite{Kiermaier:2008jy} or \cite{Kiermaier:2007jg}. 
It is still not clear why  $I_0^{(N)}$ is related to Feynman rules with such a singular gauge fixing condition. It might be because of the formulation of $I_0^{(N)}$ itself, while it is also possible to assume that it comes from the gauge fixing condition of the tachyon vacuum solution $\Psi_T$,  which is also the dressed  $\mathcal B_0$ gauge. 
It might be interesting to ask whether we can relate $I_0^{(N)}$ with a Feynman rule of more regular gauge-fixing condition
by choosing $\Psi_T$ of different gauge conditions. 
%

We also proved that the tree-level $S$-matrix calculated with the new set of Feynman rules $\mathcal R_\flat$ matches the gauge invariant quantity $I_\Psi^{(N)}$. This means  that $\mathcal R_\flat$ is valid at least for an on-shell tree-level calculation. 
As we have mentioned earlier, there is an unfinished part in the foundation of the Feynman rules $\mathcal R_\flat$, but if we could complete it, it would allow us to extend our result to the loop level. 

We believe that our analysis 
cast 
light on 
the rather unconventional object $A_T-A_\Psi$ and its relation to propagators in the conventional perturbative calculation.
It would be great if our work could help obtaining a simplified expression or description for the loop amplitudes.\\

\appendix

\section{$T_n$ and Feynman diagrams}
In this appendix, we will provide proofs of  the statements regarding $T_n$ and scattering amplitudes. 
In particular, the formula \eqref{eq:Npt_prop}, which we will give at the end of this section, 
can be regarded as 
a 
graphical proof of the equivalence of \eqref{eq:fml} and the on-shell scattering amplitude.

\subsection{Tree amplitudes in cubic open SFT}
\label{sec:proofTnA}
This subsection presents the proof of \eqref{eq:Npt_v1}. 
Before entering to the main point, let us summarize how to calculate open string scattering amplitudes using Feynman rules. 
As we stated in Section \ref{sec:ugr57u}, 
 $N$-point amplitudes are decomposed as
\begin{align}
\mathcal{A}^{(N)}=\Big[\mathcal{A}_{12...N}+\text{ $\big((N-1)!-1\big)$ terms }\Big]\,.
\end{align}
To calculate the color-ordered amplitude, we need to sum over the possible  non-crossing diagrams with the fixed cyclic ordering of the external states. 
\\

\begin{figure}[tbp]
\begin{center}
\begin{overpic}[width=11cm, 
bb=0 94 1024 584, 
]{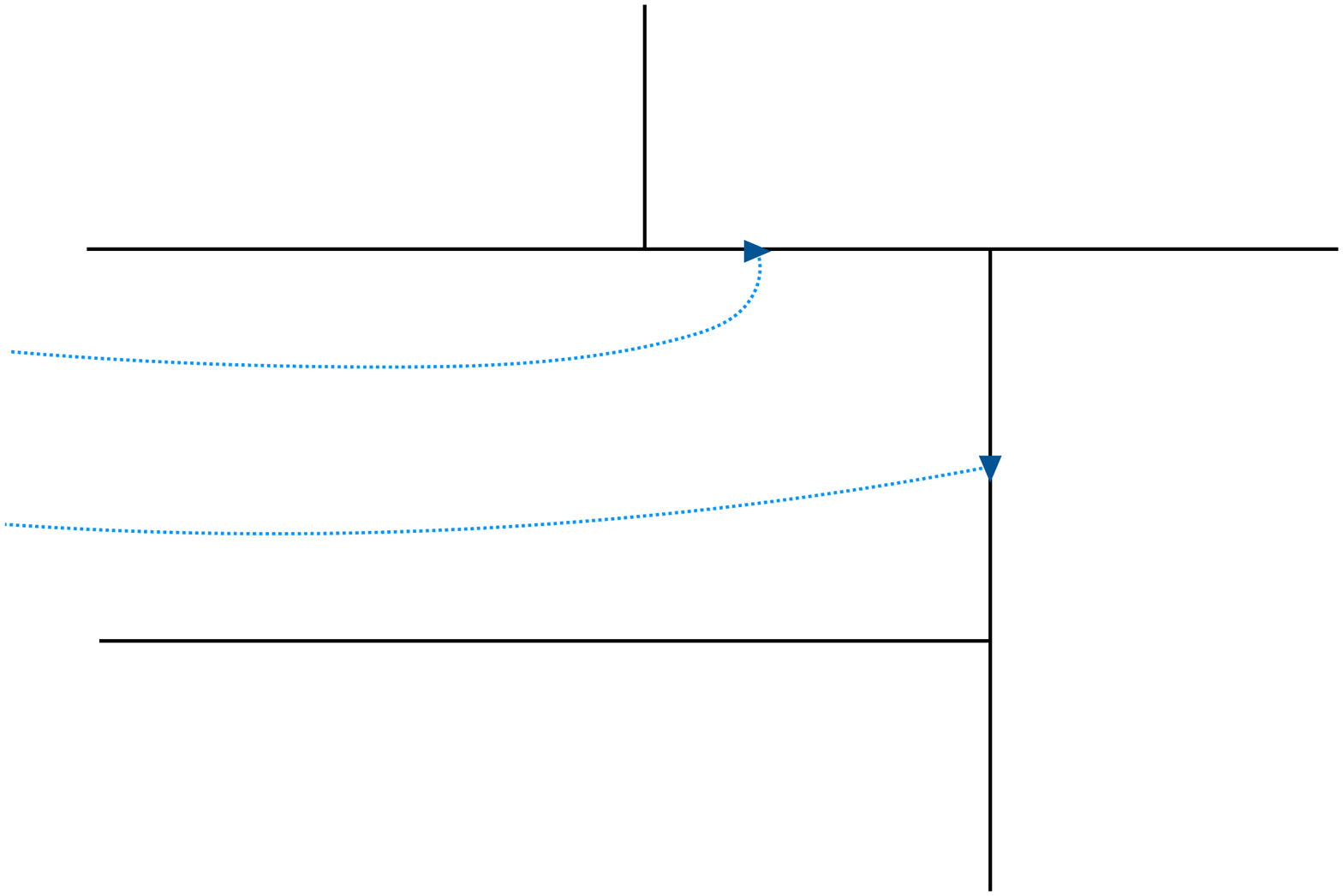}
\put(37,43){\small$\varphi_1$}
\put(8,30){\small$\varphi_2$}
\put(8,9){\small$\varphi_3$}
\put(57,0){\small$\varphi_4$}
\put(73,29){\small$\varphi_5$}
\put(5,23){\small$Y$}
\put(5,15){\small$Y$}
\end{overpic}
\caption{An illustration for the formula \eqref{eq:Npt_v1}.}
\label{figzzz}
\end{center}
\end{figure}

\noindent
\textbf{Sketch of the proof of \eqref{eq:Npt_v1}:}  
Suppose that 
we have a tree, non-crossing diagram $g$ which has $N$ external lines labeled by $\varphi_1,\varphi_2,...,\varphi_N$ in counter clockwise. 
Since $g$ is a tree diagram, we can assign each internal line (propagator) a direction away from $\varphi_1$. Let a dotted line grow from the right side of each internal line and assign it the label $Y$ as in Figure \ref{figzzz}. 
By arranging the $\varphi_j$s and $Y$s according to cyclic order in the graph, with $\varphi_1$ leftmost, we can express $g$ using the functions $Y_\text{d}$, $I$, and $m$. 
For example, the graph in Figure \ref{figzzz} corresponds to 
\begin{align}
&\varphi_1\varphi_2YY\varphi_3\varphi_4\varphi_5\notag\\
\label{eq:0245f56}
\to I(\varphi_1,&m(\varphi_2,Y_\text{d}(Y_\text{d}(\varphi_3,\varphi_4),\varphi_5))).
\end{align}
Since $m(\varphi_2, Y_\text{d}(Y_\text{d}(\varphi_3, \varphi_4),\varphi_5))$ appears in $\mathcal P^{-1}T_4(\varphi_2,\varphi_3,\varphi_4,\varphi_5)$, we see that \eqref{eq:0245f56} appears in \eqref{eq:Npt_v1} for $N=5$.  
We can confirm that this correspondence is one to one onto. $\square$

\subsection{Other expressions for $\mathcal A_{12...N}$}
\label{sec:dbfd}
In this subsection, we will present two more formulas for the scattering amplitude other than \eqref{eq:Npt_v1}.
\noindent
By focusing on propagators, we have
\begin{multline}
\label{eq:Npt_prop}
{\mathcal A}_{12... N}=\frac{1}{2(N-3)}\sum_{i=2}^{N-1}\Big[I\big(\mathcal{P}^{-1}T_i(\varphi_1,...\,,\varphi_i),T_{N-i}(\varphi_{i+1},...\,,\varphi_N)\big)\\
+\big(\text{$(N-1)$ cyclic permutations w.r.t. $\varphi_i$}\big)\Big]\,,
\end{multline}
where note that each Feynman diagram is counted $2(N-3)$ times because it contains $(N-3)$ propagators, with two possible orientations,  and so the prefactor $1/2(N-3)$ is to compensate this over-counting. \\

\begin{figure}[tbp]
\begin{center}
\begin{overpic}[width=11cm, bb=0 10 1024 668
]{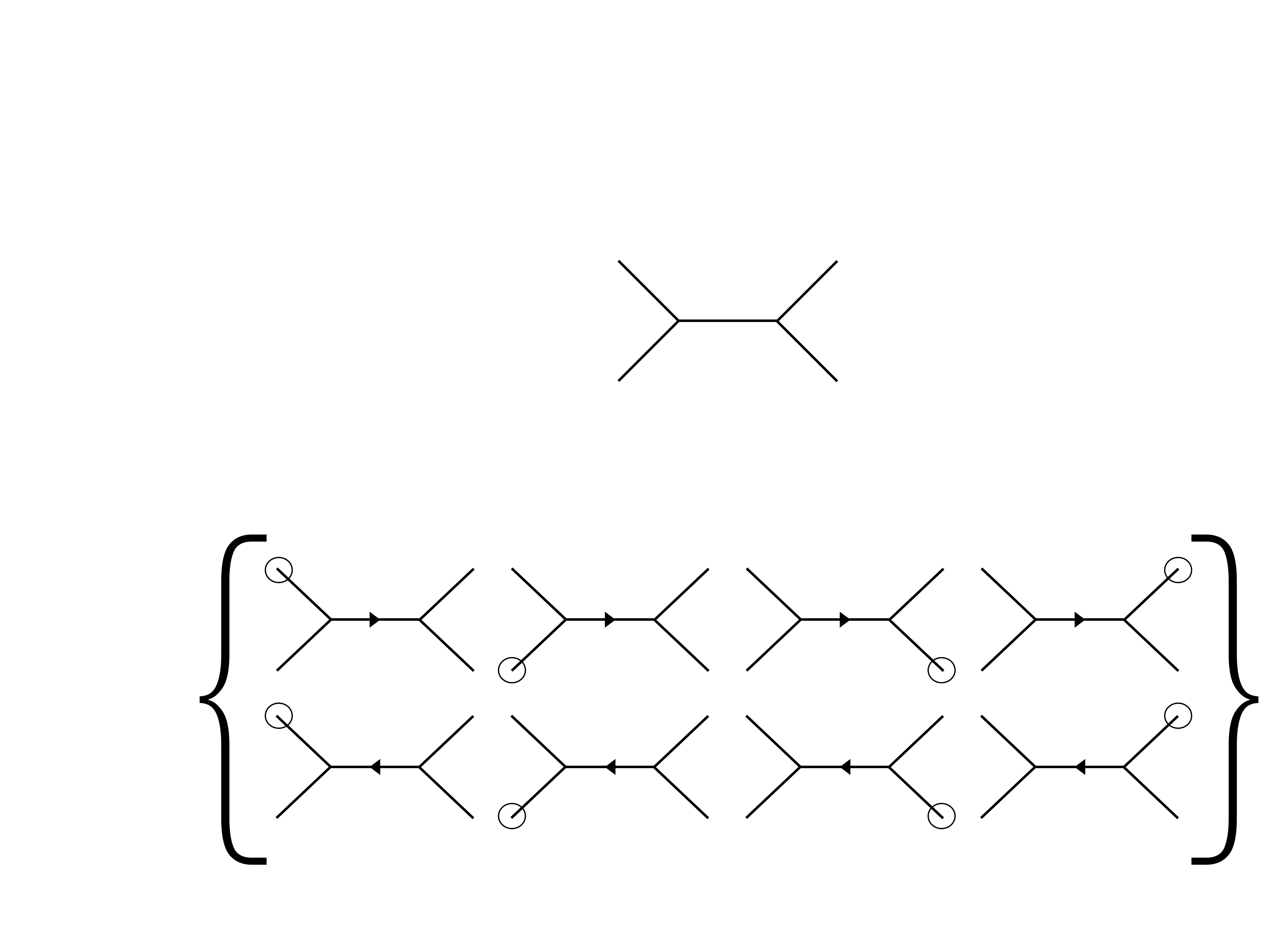}
\put(1,18){$A(g)=$}
\put(38,48){$g=
$}
\put(45,53){$\scriptsize i_1
$}
\put(45,41){$\scriptsize i_2
$}
\put(67,41){$\scriptsize i_3
$}
\put(67,53){$\scriptsize i_4
$}
\put(38,20){$\scriptsize ,$}
\put(56.7,20){$\scriptsize ,$}
\put(75.5,20){$\scriptsize ,$}
\put(93.5,20){$\scriptsize ,$}
\put(38,9){$\scriptsize ,$}
\put(56.7,9){$\scriptsize ,$}
\put(75.5,9){$\scriptsize ,$}
\end{overpic}
\caption{An example of $A(g)$.  Note that we dropped labels of the external lines on the right hand side of the second equality for simplicity.}
\label{fig:exAg}
\end{center}
\end{figure}

\noindent
\textbf{Proof of \eqref{eq:Npt_prop}:} What we 
need to prove is the following identity
\begin{equation}
\label{eq:needsproof}
2(N-3)\langle 1,N-1\rangle = N\sum_{i=2}^{N-3}\langle i, N-i\rangle.
\end{equation}
Let $G_I$ be a set of the Feynman diagrams (tree, non-crossing) with external states $\varphi_{i_1}, ... ,\varphi_{i_N}$ arranged in a counter-clockwise direction, where $I$ denotes a series of indices $I=i_1,...,i_N$. 
Let $A(g)$ denote a set of (ornamented) Feynman diagrams obtained from $g\in G_I$ by marking one of the propagators with an oriented symbol 
and marking one of the external states (Figure \ref{fig:exAg}). 
Define 
\[B_I\equiv \bigcup_{g\in G_I}A(g).\]
In the following, we will calculate the sum of Feynman diagrams in $B_I$ by ignoring the marks on propagators and external lines.  
We will see that the result is the left or right hand side of \eqref{eq:needsproof}, depending on the way of calculation.  
\textbf{Way 1:}  Let $l_1(g')$ $(g'\in B_I)$ denote the index $i$ of the marked external line of $g'$. 
First we sum over the Feynman diagrams in $B_I$ with $i=l_1(g')$ fixed. Then we sum over the index $i$. 
The result 
corresponds to the right hand side of \eqref{eq:needsproof}.\footnote
{
The factor 2 is from the orientation of the mark on propagators, and the factor $(N-3)$ is from the 
number of the 
propagators. 
} 
\textbf{Way 2:} 
By removing the marked propagator from a Feynman diagram $g'\in B_l$, 
we obtain 
an ordered pair of natural numbers $l_2(g')\equiv (i,j)$ $(i+j=N$, $\ i,\,j\ge 2)$, 
each represents the number of external states in the resulting subgraphs.  
First we sum over the Feynman diagrams in $B_I$ with a fixed $l_2(g')=(i,j)$. We then sum over $(i,j)$. 
The result 
corresponds to the left hand side of \eqref{eq:needsproof}.\footnote
{
The factor $N$ is from the position of the mark on external lines. 
}\quad $\square$ \\

\noindent
{Yet another formula for scattering amplitudes is obtained by focusing on vertices,}
\begin{align}
{\mathcal A}_{12... N}
&=\frac{1}{3(N-2)}\sum_{\substack{i+j+k=N\\[1mm] i,j,k\geq1}}
\Big[\big\langle T_i(\varphi_1,...\,,\varphi_i),T_j(\varphi_{i+1},...\,,\varphi_{i+j}),T_{k}(\varphi_{i+j+1},...\,,\varphi_N)
\big\rangle
\nonumber
\\[-5mm]
\label{eq:Npt_vertex}
&\qquad\qquad\qquad\qquad\qquad
+\big(\text{$(N-1)$ cyclic permutations w.r.t. $\varphi_i$}\big)\Big]\,,
\end{align}
where
\begin{equation}
\langle T_i, T_j, T_k\rangle=I(T_i,m(T_j,T_k)). 
\end{equation}
The formula \eqref{eq:Npt_vertex} is proved in a similar manner as in the proof of \eqref{eq:Npt_prop}; 
in this case, the factor ${1}/{(N-2)}$ cancels out the number of vertices, and the factor ${1}/{3}$ compensates for the rotation around each vertex. 
In terms of urchins, the right hand side of \eqref{eq:Npt_vertex} corresponds to that of \eqref{eq:fml}. 
Thus, derivation of the formula \eqref{eq:Npt_vertex}  provides a short alternative proof that \eqref{eq:fml} represents the $S$-matrix. 
\\

\section{Gauge invariance of  the identity \eqref{eq:correspondence1} }
\label{sec:proof}

\noindent
In this appendix, we prove the gauge invariance 
of the basic identity \eqref{eq:correspondence1} 
by considering an infinitesimal gauge transformation of it. 
The method is basically the same as that in Section 4.3 of \cite{Masuda:2019rgv}.
Let us borrow the notation from there and 
define the following quantities
\begin{equation}
H^{(p,q,r)}={\sum}''{\sum}' H_I^{(p,q,r)},
\end{equation}
\begin{equation}
T^{(p,q,r)}={\sum}''{\sum}' T_I^{(p,q,r)},
\end{equation}
\begin{equation}
{T}^{j(p,q,r)}={\sum}''{\sum}' { T}_I^{j(p,q,r)},
\end{equation}
\begin{equation}
{\tilde T}^{[j](p,q)}={\sum}''{\sum}' {\tilde T}_I^{[j](p,q)}.
\end{equation}
In this subsection we assume that all the subscripts in these expressions are elements of $\mathbb Z_N=\mathbb Z/N\mathbb Z$, where $N$ is the number of the external states. 
These quantities are all translationally invariant:
\begin{equation}
\label{eq:heishinfufhensei1234}
H^{(p, q,r)}=H^{(p+d,q+d, r+d)}, \quad T^{(p,q,r)}=T^{(p+d,q+d,r+d)},
\end{equation}
\begin{equation}
\label{eq:heishinfufhensei1234589}
T^{j(p, q,r)}=T^{j+d(p+d,q+d, r+d)}, \quad {\tilde T}^{[j](p,q)}={\tilde T}^{[j+d](p+d,q+d)}.
\end{equation}
Urchins and $H^{(p,q,r)}$ are related by\footnote{
It is more accurate to write
\begin{equation}
[q_R-p_R,r_R-q_R,N+p_R-r_R]\quad (1\le p_R<q_R<r_R\le N)
\end{equation}
for the right-hand side. 
Here $x_R\in \mathbb Z$ is the representative corresponding to $x\in {\mathbb Z}_N$. 
But we wrote it as \eqref{eq:konranshinai}, as there is no risk of confusion.
}
\begin{equation}
\label{eq:konranshinai}
H^{(p,q,r)}=[q-p,r-q,N+p-r] \quad (1< p<q<r < N).
\end{equation}
The basic relation can then be expressed as 
\begin{equation}
\sum_{\alpha=p+1}^{r-1} [\alpha-p,r-\alpha,N+p-r]= \sum_{\alpha=r+1}^{N+p-1} [r-p,\alpha-r,N+p-\alpha],
\end{equation}
or
\begin{equation}
\label{eq:7bq5b7q45b7c79btqc}
\sum_{\alpha=p+1}^{r-1}H^{(p,\alpha,r)}=\sum_{\alpha=r+1}^{N+p-1} H^{(p,r,\alpha)}.
\end{equation}
Note that this equals $[N+p-r,r-p]$. 
An infinitesimal gauge transformation of $H^{(p,q,r)}$ is
\begin{equation}
\delta H^{(p,q,r)}=\sum_{i \notin\{p,q,r\}} T^{i(p,q,r)}+{\tilde T}^{[p](q,r)}+{\tilde T}^{[q](p,r)}+{\tilde T}^{[r](p,q)}. 
\end{equation}
It holds that
\begin{equation}
{\tilde T}^{[i](p,q)}=\sum_{r\neq p,q,i}s(r,i;p,q)T^{i(p,q,r)}. 
\end{equation}
Note that the sign factor $s(w,x;y,z)$ is well-defined even for its arguments are elements of $\mathbb Z_N$. 
Now, let us consider an infinitesimal gauge transformation of each side of \eqref{eq:7bq5b7q45b7c79btqc}:
\begin{equation}
\begin{split}
\delta \text{(l.h.s.)}=&\sum_{\alpha=p+1}^{r-1}\delta H^{(p,\alpha,r)} \\
=&\sum_{\alpha=p+1}^{r-1}\Bigg( \sum_{i\notin \{p,\alpha,r \}}T^{i(p,\alpha,r)}
+\sum_{x\ne p,\alpha, r}s(x,p;\alpha,r)T^{p(\alpha,r,x)} \\
&\qquad\qquad +\sum_{x\ne p,\alpha, r}s(x,\alpha;p,r)T^{\alpha(p,r,x)} 
+\sum_{x\ne p,\alpha, r}s(x,r;p,\alpha)T^{r(p,\alpha,x)} \Bigg)\\
=&A_L+B_L
\end{split}
\end{equation}
where
\begin{equation}
A_L=\sum_{\alpha=p+1}^{r-1}\sum_{i=r+1}^{N+p-1}T^{i(p,\alpha,r)}
+\sum_{j=p+1}^{r-1}\sum_{x=r+1}^{N+p-1}T^{j(p,x,r)},
\end{equation}
\begin{equation}
B_L=\sum_{\alpha=p+1}^{r-1}\sum_{x\ne p,\alpha, r}s(x,p;\alpha,r)T^{p(\alpha,r,x)}
+\sum_{\alpha=p+1}^{r-1}\sum_{x\ne p,\alpha, r}s(x,r;p,\alpha)T^{r(p,\alpha,x)}.
\end{equation}
Whereas
\begin{equation}
\begin{split}
\delta \text{(r.h.s.)}=&\sum_{\alpha=r+1}^{N+p-1}\delta H^{(p,\alpha,r)} \\
=&\sum_{\alpha=r+1}^{N+p-1}\Bigg( \sum_{i\notin \{p,\alpha,r \}}T^{i(p,\alpha,r)}
+\sum_{x\ne p,\alpha, r}s(x,p;\alpha,r)T^{p(\alpha,r,x)} \\
&\qquad\qquad +\sum_{x\ne p,\alpha, r}s(x,\alpha;p,r)T^{\alpha(p,r,x)} 
+\sum_{x\ne p,\alpha, r}s(x,r;p,\alpha)T^{r(p,\alpha,x)} \Bigg)\\
=&A_R+B_R,
\end{split}
\end{equation}
where
\begin{equation}
A_R=\sum_{\alpha=r+1}^{N+p-1}\sum_{i=p+1}^{r-1}T^{i(p,\alpha,r)}
+\sum_{j=r+1}^{N+p-1}\sum_{x=p+1}^{r-1}T^{j(p,x,r)}
\end{equation}
\begin{equation}
B_R=\sum_{\alpha=r+1}^{N+p-1}\sum_{x\ne p,\alpha, r}s(x,p;\alpha,r)T^{p(\alpha,r,x)}
+\sum_{\alpha=r+1}^{N+p-1}\sum_{x\ne p,\alpha, r}s(x,r;p,\alpha)T^{r(p,\alpha,x)}.
\end{equation}
It holds that
\begin{equation}
A_L=A_R, \ 
\text{and }
B_L=B_R. 
\end{equation}
Thus we conclude that \eqref{eq:7bq5b7q45b7c79btqc} holds regardless of gauge fixing condition of $\Psi$ or $\Psi_T$. \quad $\square$

\section{
More on unconventional propagators
}
In Appendix \ref{sec:c1}, we present some propagators other than \eqref{eq:64ryfh}, with which the Feynman diagrams calculated is given by sum of urchins. 
In particular, the simplified propagator given in \eqref{eq:hh3hy7d} does not satisfy the BPZ property but can be used to 
calculate on-shell scattering amplitudes as in the case of $\mathcal P_\text{s}$.
In Appendix \ref{sec:c2}, we first present a formal argument that the on-shell amplitude does not depend on the choice of propagators. 
Then, we show that this argument fails when we consider the unconventional propagator ${\mathcal P}_\flat$. 

\subsection{Other unconventional propagators}
\label{sec:c1}
Note that our propagator ${\mathcal P}_\flat$ can be written as 
\begin{equation}
{\mathcal P}_\flat=\frac{1}{2}(\mathcal P_\triangle+\mathcal P_\triangle^\star)
\end{equation}
where the simplified propagator $\mathcal P_\triangle$ and its BPZ conjugation are given by
\begin{equation}
\label{eq:hh3hy7d}
\mathcal P_\triangle \phi= \frac{A}{W_\Psi}\phi, \quad \mathcal P_\triangle^\star \phi=(-)^{\mathrm{gh}(\phi)} \phi\frac{A}{W_\Psi}. 
\end{equation}
This simplified propagator satisfies $\{Q, \mathcal P_\triangle\}=1$, 
and can be used to calculate on-shell amplitude if we are careful about its orientation, just for the same reason that \eqref{eq:propagator2} can be used for calculation of the $S$-matrix.
%

In addition to $\mathcal P_\flat$, it is natural to consider 
a propagator 
\begin{equation}
{\mathcal P}_\sharp=
{\mathcal P}_\triangle Q  {\mathcal P}^{\star}_\triangle,
\end{equation}
which is 
the other combination of $\mathcal P_\triangle$ and $\mathcal P_\triangle^\star$ respecting the BPZ property.   
We can check immediately that this second propagator gives a result consistent with $I^{(N)}_\Psi$.
Indeed, we have
\begin{align}
T^\sharp_n(\varphi_1,\ldots,\varphi_n)&=(-1)^n\frac{1}{\sqrt{-W}}A\mathcal{O}_1A\mathcal{O}_2\cdots A\mathcal{O}_{n-1}W\mathcal{O}_n\sqrt{-W}\,.
\end{align}
Here $T^\sharp_n$ is defined by \eqref{eq:defT00} and \eqref{eq:defT} with the propagator replaced by ${\mathcal P}_\sharp$. 
From \eqref{eq:Npt_v1}, we conclude that 
\begin{equation}
\begin{split}
\mathcal{A}_{12... N}=I\big(T^\sharp_1(\varphi_1),\mathcal{P}^{-1}T^\sharp_{N-1}(\varphi_2,...\,,\varphi_N)\big).
\end{split}
\end{equation}

\subsection{$\mathcal P$-independence of on-shell amplitudes: A conventional argument }
\label{sec:c2}

Let us pretend that we have two operators, $\mathcal{P}_1$ and $\mathcal{P}_2$, which satisfy
\begin{align}
\{Q,\mathcal{P}_1\}=\{Q,\mathcal{P}_2\}=1\,,
\quad
\mathcal{P}_1^\star=\mathcal{P}_1\,,
\quad
\mathcal{P}_2^\star=\mathcal{P}_2\,.
\end{align}
We wish to show that scattering amplitudes computed with them are identical.
For this purpose, let us define the following one-parameter family of operators:
\begin{align}
\mathcal{P}(x)=x\mathcal{P}_1+(1-x)\mathcal{P}_2\,,
\end{align}
which satisfies
\begin{align}
\{ Q,\mathcal{P}(x)\}=1\,,
\quad
\mathcal{P}(x)^\star=\mathcal{P}(x)\,.
\end{align}
A useful property is that its derivative in $x$ is BRST exact:
\begin{align}
\partial_x\mathcal{P}(x)=\mathcal{P}_1-\mathcal{P}_2=\{Q,\mathcal{P}_2\}\mathcal{P}_1-\mathcal{P}_2\{Q,\mathcal{P}_1\}=[Q,\mathcal{P}_2\mathcal{P}_1]\,.
\end{align}
The $x$-derivative of scattering amplitudes computed with $\mathcal{P}=\mathcal{P}(x)$ reads
\begin{align}
\partial_xA_{12... N}&=\sum_{i=2}^{N-1}I\big(\mathcal{P}^{-1}T_i(\varphi_1,...\,,\varphi_i),(\partial_x\mathcal{P})\mathcal{P}^{-1}T_{N-i}(\varphi_{i+1},...\,,\varphi_N)\big)
\nonumber
\\
&=\sum_{i=2}^{N-1}I\big(\mathcal{P}^{-1}T_i(\varphi_1,...\,,\varphi_i),[Q,\mathcal{P}_2\mathcal{P}_1]\mathcal{P}^{-1}T_{N-i}(\varphi_{i+1},...\,,\varphi_N)\big)
\nonumber
\\
\label{eq:u3hf1}
&=0\,,
\end{align}
where we used \eqref{eq:cdyei3n}  at the last equality. 
(Also, the first equality follows by the same reasoning as in the proof of \eqref{eq:Npt_prop} in Appendix \ref{sec:dbfd}.)
Since the amplitudes are $x$-independent, those computed with $\mathcal{P}_1$ and $\mathcal{P}_2$ are identical and so amplitudes are independent of the choice of the propagator. Here notice that we have shown that tree-level amplitudes are invariant under replacement of the propagator by an arbitrary operator $\mathcal{P}$ satisfying $\{Q,\mathcal{P}\}=1$ and $\mathcal{P}^\star=\mathcal{P}$.\footnote{
The discussion in this subsection is a simplified version of that in \cite{Masuda:2019_2}, 
and 
by inserting projection operators regarding domains of $\mathcal{P}_i$s into formulas in this subsection, we can reproduce the formal argument in \cite{Masuda:2019_2}. 
}

Now, we try applying this argument to the propagator ${\mathcal P}_\flat$. 
Suppose 
$\mathcal P_1={\mathcal P}_\flat$ and $\mathcal P_2={\mathcal P}_\sharp$. 
In \eqref{eq:u3hf1}, we have assumed that a state of the form 
\begin{equation}
\label{eq:x6lf4f}
Q
{\mathcal P}_\sharp {\mathcal P}_\flat\phi
\end{equation} 
is BRST exact and its contraction with the BRST-closed state is zero. However, this does not hold in the present case. 
To see this, let us consider the following state:
\begin{equation}
Q{\mathcal P}_\sharp {\mathcal P}_\flat \phi_{123}
=\frac{B}{K}\phi_1\ast\phi_2\ast\phi_3+\phi_1\ast\phi_2\ast\phi_3 \frac{B}{K}, 
\end{equation}
where $\phi_{123}=\phi_1\ast\phi_2\ast\phi_3$ with $\phi_j$ an element of the BRST cohomology at ghost number 1. 
If this state was BRST exact, the contraction of it with $\phi_4$ would vanish.  
In fact, the result is not zero but an integration over part of the moduli space
\begin{multline}
\label{eq:83bdf6crswgw}
\int \phi_4 Q
{\mathcal P}_\sharp{\mathcal P}_\flat
\left(\frac{B}{K}\phi_1\ast\phi_2\ast\phi_3+\phi_1\ast\phi_2\ast\phi_3 \frac{B}{K}\right)\\
=
\int_0^1 dx\langle \phi_3(0)\phi_4(x)\phi_1(1)\phi_2(\infty)\rangle_\text{UHP}, 
\end{multline}
as calculated in Section 5 of \cite{Masuda:2019rgv}.\footnote{The section number is based on the first version of the preprint, arXiv:1908.09784v1 [hep-th]. 
} 
This expression \eqref{eq:83bdf6crswgw} is reminiscent of the formal expression  for $S$-matrix as the Witten integral of a BRST-exact state, which is presented in Section 4.4 of \cite{Masuda:2019rgv}.

The formal proof above is thus not satisfactory, if not broken,  due to the problem caused by $B/K$. 
This is why we calculated the sum of the Feynman diagrams to prove the validity of $\mathcal R_\flat$ for the on-shell amplitudes in Section~\ref{sec:ddggdfdlllp3}. 

\noindent
\subsubsection*{Acknowledgement} 
The authors would like to thank the organizers of  the online workshop, ``Workshop on Fundamental Aspects of String Theory  and Related Aspects" (ICTP-SAIFR/IFT-UNESP, June 1-12, 2020),
and participants of the workshop for valuable discussion and comments. 
This research of TM has been supported by the GACR grant 20-25775X, while the research of HM by the GACR grant 19-28268X. 
T.N. is supported in part by JSPS KAKENHI Grant Numbers JP17H02894 and JP20H01902.

\noindent

\clearpage

\end{document}